  \documentstyle[aas2pp4,epsf]{article}

\newcommand{\Nco}{N_{\rm e}^*} 

\lefthead{Hirotani}
\righthead{}

\begin{document}

\title{KINETIC LUMINOSITY AND COMPOSITION OF
       ACTIVE GALACTIC NUCLEI JETS}
 \author{Kouichi Hirotani}
  \affil{Max-Planck-Institut f$\ddot{\rm u}$r Kernphysik,
        Postfach 103980, 
	D-69029 Heidelberg, Germany\\
        hirotani@mpi-hd.mpg.de}

\begin{abstract}
We present a new method how to discriminate 
the matter content of parsec-scale jets of active galactic nuclei.
By constraining the kinetic luminosity of a jet
from the observed core size at a single
very long baseline interferometry frequency,
we can infer the electron density
of a radio-emitting component as a function of the composition.
Comparing this density with that obtained from 
the theory of synchrotron self-absorption,
we can determine the composition.
We apply this procedure to the five components in the 3C~345 jet
and find that they are likely pair-plasma dominated at 11 epochs out of
the total 21 epochs,
provided that the bulk Lorentz factor is less than 15
throughout the jet.
We also investigate the composition of the 3C~279 jet
and demonstrate that its two components are likely pair-plasma dominated
at three epochs out of four epochs,
provided that their Doppler factors are less than 10,
which are consistent with observations.
The conclusions do not depend on the lower cutoff energy of
radiating particles.
\end{abstract}



\section{Introduction}
\label{sec:intro}

The study of extragalactic jets on parsec scales is
astrophysically interesting in the context of
the activities of the central engines of
active galactic nuclei (AGN).
In particular, a determination of their matter content
would be an important step in the study of jet formation,
propagation and emission.
There are two main candidates for their matter content:
A \lq normal plasma' consisting of (relativistic or non-relativistic)
protons and relativistic electrons
(Celotti and Fabian 1993; see also
 G$\acute{\rm o}$mez et al. 1993, 1994a,b
 for numerical simulations of shock fronts in such a jet)
and a \lq pair plasma' consisting only of relativistic electrons
and positrons (e.g., Kino and Takahara 2004).
Discriminating between these possibilities is crucial for
understanding the physical processes occurring close to the
central engine (presumably a supermassive black hole)
in the nucleus.

Very long baseline interferometry (VLBI)
is uniquely suited to the study of the matter content of
pc-scale jets,
because other observational techniques cannot image
at milliarcsecond resolution and must resort to indirect means of
studying the active nucleus.
So far, there have been two observational approaches with VLBI 
to discriminate jet compositions: 
polarization (Wardle et al. 1998) and
spectroscopic (Reynolds et al. 1996) techniques.

Wardle et al. (1998) observed the 3C~279 jet at 15 and 22 GHz.
They examined the {\it circular} polarization 
($\sim 1$\%) due to Faraday {\it conversion} 
of linear to circular polarization caused by low-energy electrons 
and derived that the lowest Lorentz factor, $\gamma_{\rm min}$,
of radiating particles, should be much less than $10^2$.
They also analyzed the {\it linear} polarization ($\sim 10$\%),
which strongly limits internal Faraday {\it rotation}
to derive that $\gamma_{\rm min} \gg 10^2$ must be held for a
normal-plasma composition 
and that no constraint would be obtained for a pair-plasma composition.
(Note that an equal mixture of electrons and positrons can produce
Faraday conversion, but not rotation.)
From these arguments, they concluded that component CW of the
3C~279 jet consists of a pair plasma with $\gamma_{\rm min} \ll 10^2$. 

\begin{figure}
\centerline{ \epsfxsize=8cm \epsfbox[100 120 500 350]{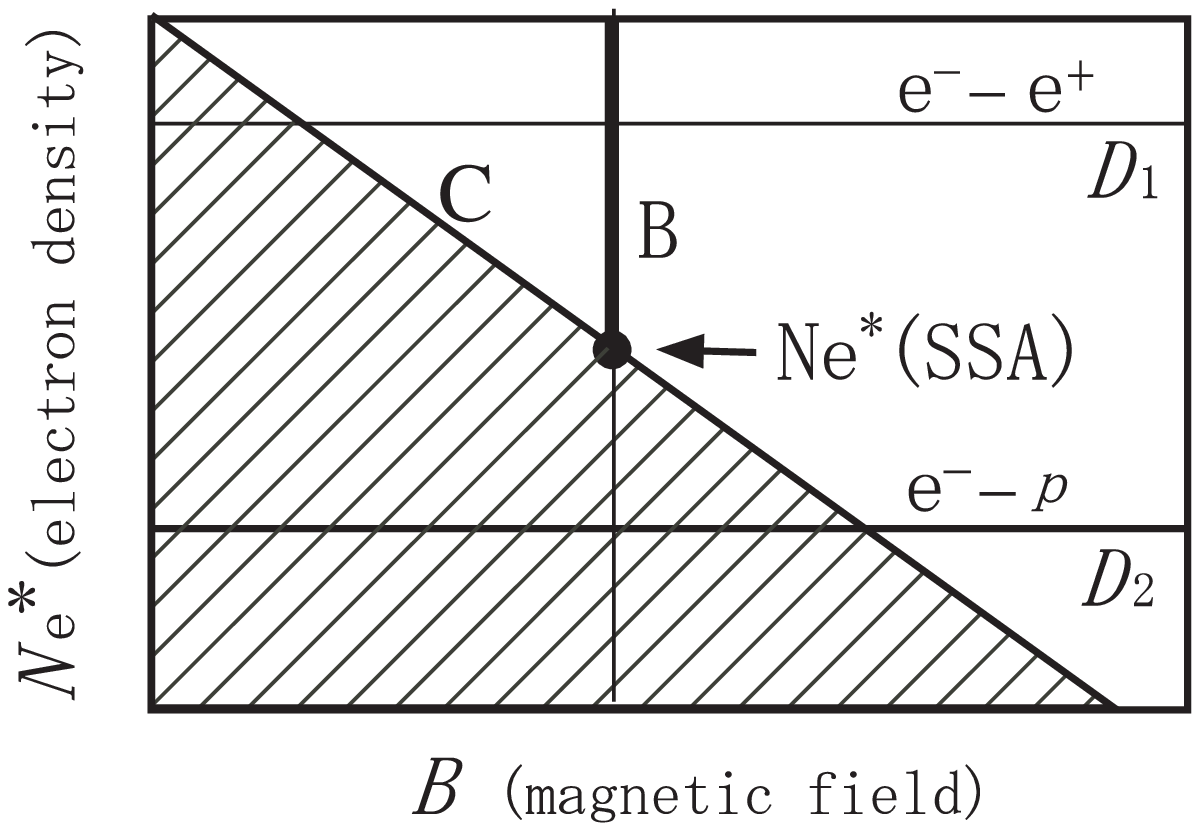} }
\caption{\label{fig:NeB}
Constraints on ($B$,$\Nco$) plane
imposed by synchrotron self-absorption and
total kinetic luminosity considerations
(see fig.~1 in Reynolds et al. 1996).
The VLBI surface brightness constraint on $B$ gives line B,
while the condition that the a jet component
(or the resolved core in Reynolds et al. 1996)
becomes optically thick to self-absorption gives line C.
The horizontal lines are the proper densities
derived from the kinetic luminosity;
line ${\rm D}_1$ and ${\rm D}_2$ correspond to
a pure pair and a pure normal plasma, respectively.
        }
\end{figure}

On the other hand, 
Reynolds et al. (1996) analyzed historical VLBI data
of the M87 jet at 5 GHz (Pauliny-Toth et al. 1981)
and concluded that the core is probably dominated
by an $e^\pm$ plasma.
In the analysis, they utilized the standard theory of synchrotron
self absorption (SSA) to constrain the magnetic field, $B$ [G],
and gave $B=0.2$~G (line B in figure~\ref{fig:NeB}).
They also considered the condition that the resolved core
becomes optically thick for self-absorption
and derived another constraint on $\Nco$ and $B$.
Using the observed angular diameter of the core ($0.7$mas),
the total core flux density ($1.0$~Jy) at 5~GHz, and
the reported viewing angle ($30^\circ - 40^\circ$),
and assuming that the spectral index is $0.5$ (see end of this section),
they derived
$\Nco B^2 > 2 \delta_{\rm max}^{-1}$
in cgs unit (line C in figure~\ref{fig:NeB}),
where
$\delta_{\max}$ ($=2$ for the M87 core)
refers to the upper limit of the
Doppler factor of the fluid's bulk motion.

This condition is, however, applicable only for the VLBI observations
of M87 core at epochs September 1972 and March 1973.
Therefore, in order to apply the analogous method to other AGN jets
or to the M87 jet at other epochs,
we must derive a more general condition.

On these grounds, Hirotani et al. (1999, hereafter Paper~I)
generalized the condition
$ \Nco B^2 > 2 \delta_{\rm max}^{-2} $
and applied it to the 3C~279 jet on parsec scales.
In that paper, they revealed that the core and components C3 and C4,
of which spectra are reported,
are likely dominated by a pair plasma.
It is interesting to note that the same conclusion
was derived by an independent method by Wardle et al. (1998)
on component CW in the 3C~279.

Subsequently, Hirotani et al. (2000, hereafter Paper~II)
applied the same method to the 3C~345 jet.
Deducing the kinetic luminosity, $L_{\rm kin}$,
from a reported core-position offset (Lobanov 1998),
they demonstrated that components C2, C3, and C4 at epoch 1982.0,
C5 at 1990.55, and C7 at four epochs are likely 
dominated by a pair plasma.
In the present paper, 
we propose a new scheme to deduce the kinetic luminosity
of an unresolved core and 
re-examine the composition of the two blazers 3C~345 and 3C~279,
adding spectral information at 11 epochs for the 3C~345 jet
to Paper~II and and 2 epochs for the 3C~279 jet to Paper~I.

In the next section,
we give a general expression of lines B and C in figure~\ref{fig:NeB}.
We then describe in \S~\ref{sec:lineD} 
a new method how to infer $L_{\rm kin}$
from the core size observed at a single VLBI frequency;
the inferred $L_{\rm kin}$ gives lines~${\rm D}_1$ and ${\rm D}_2$
in figure~\ref{fig:NeB},
depending on the composition.
Once $L_{\rm kin}$ is obtained for the core by this method,
we can apply it to individual jet components,
assuming a constant $L_{\rm kin}$ along the jet.
In sections \ref{sec:3C345_appli} and \ref{sec:3C279_appli},
we examine the two blazers 3C~345 and 3C~279.
In the final section,
we discuss the possibility of the entrainment of the ambient matter
as a jet propagates downstream.

We use a Hubble constant $H_0 = 65 h$ km/s/Mpc and $q_0 =0.5$
throughout this paper.
Spectral index $\alpha$ is defined
such that $S_\nu \propto \nu^{+\alpha}$.

\section{Synchrotron self-absorption constraint in the jet region}
\label{sec:SSA}
In this paper, we model a jet component
as a homogeneous sphere of angular diameter $\theta_{\rm d}$,
containing a tangled magnetic field $B$ [G]
and relativistic electrons, 
which give a synchrotron spectrum with
optically thin index $\alpha$.
In \S~\ref{sec:lineB},
we constrain $B$ (i.e., line~B in fig.~\ref{fig:NeB}) 
for arbitrary optical thickness
by the theory of synchrotron self-absorption.
Once $B$ is obtained, we can compute the electron number density
$\Nco({\rm SSA})$ (i.e., line~C)
as will be described in \S~\ref{sec:lineC}
with the aid of the computed optical depth at the turnover frequency. 

\subsection{Magnetic field strength}
\label{sec:lineB}

The magnetic field strength, $B$, can be computed from
the synchrotron spectrum,
which is characterized by the peak frequency, $\nu_{\rm m}$,
the peak flux density, $S_{\rm m}$,
and the spectral index, $\alpha$, in the optically thin regime.
Marscher (1983) considered a uniform spherical synchrotron source
and related $B$ with $\nu_{\rm m}$, $S_{\rm m}$, and $\alpha$
for optically thin cases ($0 \geq \alpha \geq -1.0$).
Later, Cohen (1985) considered a uniform slab of plasma
as the synchrotron source and derived smaller values of $B$
for the same set of ($\nu$,$S_{\rm m}$,$\alpha$).
In this section, we express $B$ (i.e., line~B in fig.~\ref{fig:NeB}) 
in terms of ($\nu$,$S_{\rm m}$,$\alpha$)
for a uniform sphere for {\it arbitrary} optical thickness
(i.e., for arbitrary $\alpha \leq 0$).
We assume that the magnetic field is uniform
and that the distribution of radiating particles
are uniform in the configuration space
and are isotropic and power-law (eq.[\ref{eq:ele-dist}])
in the momentum space.

\begin{figure}
\centerline{ \epsfxsize=8cm \epsfbox[0 0 500 100]{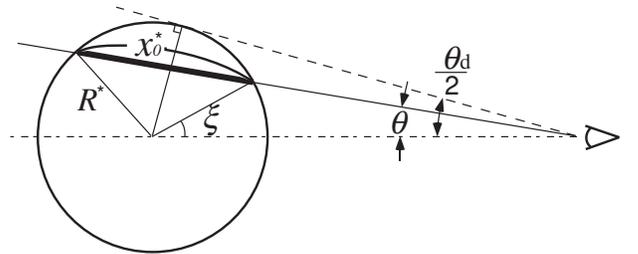} }
\caption{\label{fig:thickness}
Flux emitted from a spherical cloud of radius $R^\ast$.
        }
\end{figure}

For a uniform $B$ and $\Nco$, the transfer equation
gives the specific intensity
\begin{equation}
  I_\nu{}^\ast
    = A \nu^\ast{}^{5/2}
      \left[ 1- \exp(-\alpha_\nu{}^\ast x_0{}^\ast) \right],
  \label{eq:intensity_1}
\end{equation}
where
\begin{equation}
  A(\alpha)
  \equiv \left( \frac32 \right)^{-\alpha}
         \frac{e}{c}
         \frac{a(\alpha)}{C(\alpha)}
         \left( \frac{e}{2\pi m_{\rm e} c} \right)^{-3/2}
         B^{-1/2},
  \label{eq:def_A}
\end{equation}
and $x_0{}^\ast$ gives the physical thickness of the emitting region
along the line of sight;
$e$, $m_{\rm e}$, and $c$ refer to the charge on an electron,
the rest mass of an electron, and the speed of light, respectively.
The coefficients $a(\alpha)$ and $C(\alpha)$
are given in table~1;
a quantity with an asterisk is measured in the co-moving frame,
while that without an asterisk in the observer's frame.
At the distance $\theta$ away from the cloud center 
(fig.~\ref{fig:thickness}),
the fractional thickness, which is Lorentz invariant, becomes
\begin{equation}
  \frac{x_0{}^\ast}{2 R^\ast}
  = \cos(\theta+\xi)
  = \sqrt{ 1 -\left[\frac{\sin\theta}{\sin(\theta_{\rm d}/2)}\right]^2 },
  \label{eq:frac_thick}
\end{equation}
where 
$R^\ast\cos\xi+(R^\ast \sin\xi/\sin\theta)\cos\theta
 =R^\ast/\sin(\theta_{\rm d}/2)$
is used in the second equality;
$\theta_{\rm d}$ is the angular diameter of the component
in the perpendicular direction of the jet propagation.
Thus, the specific intensity at angle $\theta$ away from the center, 
becomes
\begin{eqnarray}
  \lefteqn{I_\nu(\theta)
           = \left( \frac{\delta}{1+z} \right)^{1/2}
             A \nu^{5/2}}
  \nonumber\\
  & & \hspace*{-0.5 truecm}\times
  \left\{ 1-\exp\left[ -\tau_\nu(0)
             \sqrt{ 1 -\left[\frac{\sin\theta}{\sin(\theta_{\rm d}/2)}
                       \right]^2 }\,
                 \right]
    \right\},
  \label{eq:intensity_2}
\end{eqnarray}
where $\tau_\nu(0) \equiv \alpha_\nu{}^\ast \cdot 2R^\ast$ is the
optical depth for $\theta=0$ (or $b=0$).
Averaging over pitch angles of the isotropic electron power-law
distribution (eq. [\ref{eq:ele-dist}]),
we can write down the absorption coefficient 
in the co-moving frame as
(Le Roux 1961, Ginzburg \& Syrovatskii 1965)
\begin{equation}
  \alpha_\nu^* = C(\alpha) r_\circ{}^2 k_{\rm e}^\ast
                 \frac{\nu_\circ}{\nu^*}
\left( \frac{\nu_{\rm B}}{\nu^*}
                 \right)^{(-2\alpha+3)/2} ,
  \label{eq:abs-coeff}
\end{equation}
where $\nu_\circ \equiv c/r_\circ \equiv c/[e^2 / (m_{\rm e} c^2)]
= 1.063 \times 10^{14}$ GHz
and $\nu_{\rm B} \equiv eB / (2\pi m_{\rm e}c)$.
In deriving equation~(\ref{eq:intensity_2}),
we utilized $I_\nu / \nu^3 = I_\nu{}^\ast / \nu^\ast{}^3$ and
$\nu/\nu_\ast=\delta/(1+z)$,
where $z$ represents the redshift and 
the Doppler factor, $\delta$, is defined by
\begin{equation}
  \delta \equiv \frac{1}{\Gamma(1-\beta\cos\varphi)},
  \label{eq:def_delta}
\end{equation}
where $\Gamma \equiv 1 / \sqrt{1-\beta^2}$ is the bulk Lorentz factor
of the jet component moving with velocity $\beta c$,
and $\varphi$ refers to the viewing angle.

Even if special relativistic effects are important,
the observer find the shape to be circular.
We thus integrate $I_\nu(\theta)\cos\theta$ over
the emitting solid angles
$2\pi \sin\theta d\theta$ in $0 \leq \theta \leq \theta_{\rm d}/2$
to obtain the flux density
\begin{eqnarray}
  S_\nu &=& 2\pi \int_0^{\theta_{\rm d}/2}
                   I_\nu(\theta) \cos\theta\sin\theta d\theta
  \nonumber \\
        & & \hspace*{-1.2 truecm}
          = \pi \sin^2\left( \frac{\theta_{\rm d}}{2} \right)\cdot
            \left(\frac{\delta}{1+z}\right)^{1/2} 
  \nonumber \\
        & & \hspace*{-0.9 truecm}
            \times A \nu^{5/2}
            \int_0^1 \left( 1-{\rm e}^{-\tau_\nu(0)\sqrt{1-\chi}}\right)
            d\chi,
  \label{eq:flux_density}
\end{eqnarray}
where $\chi \equiv [\sin\theta/\sin(\theta_{\rm d}/2)]^2$.

Differentiating equation~(\ref{eq:flux_density})
with respect to $\nu$, and
putting $dS_\nu/d\nu$ to be $0$, we obtain
the equation that relates $\tau_\nu(0)$ and $\alpha$
at the turnover frequency, $\nu_{\rm m}$,
\begin{eqnarray}
  &&
  \int_0^1 \left[ 1-{\rm e}^{-\tau_\nu(0)\sqrt{1-\chi}} \right] d\chi
  \nonumber\\
  &=& \left( 1-\frac{2}{5}\alpha \right)
      \int_0^1 \tau_\nu(0)\sqrt{1-\chi}
               {\rm e}^{-\tau_\nu(0)\sqrt{1-\chi}} d\chi.
  \label{eq:opt_depth} 
\end{eqnarray}
We denote the solution $\tau_\nu(0)$ at $\nu=\nu_{\rm m}$ as
$\tau_{\rm m}(0)$,
which is presented as a function of $\alpha$ in table~1.

It is worth comparing equation~(\ref{eq:flux_density}) with
that for a uniform slab of plasma with physical thickness $R^\ast$.
If the slab extends over a solid angle
$\pi(0.5\theta_{\rm d}/{\rm rad})^2$,
the flux density becomes
\begin{equation}
  S_\nu = \frac{\pi}{4} \left( \frac{\theta_{\rm d}}{\rm rad} \right)^2
          \left(\frac{\delta}{1+z}\right)^{1/2} A \nu^{5/2}
          \left( 1-{\rm e}^{-\tau_\nu(0)} \right).
  \label{eq:flux_density_slab}
\end{equation}
The optical depth $\tau_\nu(0)= \alpha_\nu^\ast \cdot 2R^\ast$
does not depend on $\theta$ for the slab geometry.
Therefore, comparing equations~(\ref{eq:flux_density}) and
(\ref{eq:flux_density_slab})
we can define the effective optical depth, $\langle \tau_\nu \rangle$,
for a uniform sphere of plasma as
\begin{equation}
  {\rm e}^{-\langle \tau_\nu \rangle}
  \equiv \int_0^1 {\rm e}^{-\tau_\nu(0)\sqrt{1-\chi}} d\chi.
  \label{eq:mean_opt_depth}
\end{equation}
We denote $\langle\tau_\nu\rangle$ at $\nu=\nu_{\rm m}$ as
$\langle\tau_{\rm m}\rangle$,
which is presented in table~1.
Note that $\langle\tau_{\rm m}\rangle$ is less than
$\tau_{\rm m}(0)$ due to the geometrical factor $\sqrt{1-\chi}$.

Using $\tau_{\rm m}(0)$, we can evaluate the peak flux density,
$S_{\rm m}$, at $\nu=\nu_{\rm m}$
and inversely solve equation~(\ref{eq:flux_density}) for $B$.
We thus obtain

\begin{equation}
  B = 10^{-5} b(\alpha)
      \left( \frac{\nu_{\rm m}}{\rm GHz} \right)^5
      \left( \frac{\theta_{\rm d}}{\rm mas} \right)^4
      \left( \frac{     S_{\rm m}}{\rm Jy}  \right)^{-2}
      \frac{\delta}{1+z},
  \label{eq:lineB}
\end{equation}
where $\theta_{\rm d} \ll 1 \,\mbox{rad}$ is used and
\begin{eqnarray}
  b(\alpha)
  &=& 3.98 \times 10^3 \left( \frac32 \right)^{-2\alpha}
      \left[ \frac{a(\alpha)}{C(\alpha)} \right]^2
  \nonumber\\
  \qquad
  &\times&
  \left\{ \int_0^1 \left[ 1-{\rm e}^{-\tau_{\rm m}(0)\sqrt{1-\chi}}
                   \right] d\chi
  \right\}^2.
  \label{eq:def_b}
\end{eqnarray}
The values are tabulated in table~1;
they are, in fact, close to the values obtained by Cohen (1985),
who presented $b(\alpha)$ for a slab geometry with $-\alpha \leq 1.25$.
This is because the averaged optical depth,
$\langle\tau_\nu\rangle$ at $\nu=\nu_{\rm m}$,
for a spherical geometry
becomes comparable with that for a slab geometry
(Scott \& Readhead 1977).

We could expand the integrant in equation~(\ref{eq:flux_density})
in the optically thin limit $\tau_{\rm m}(0) \ll 1$ as
\begin{equation}
  1-{\rm e}^{-\tau_{\rm m}(0)\sqrt{1-\chi}}
  \approx \tau_{\rm m}(0)\sqrt{1-\chi},
  \label{eq:opt_depth_2}
\end{equation}
to obtain $b(-0.5)= 3.34$ and $b(-1.0)= 3.85$, for example.
This optically thin limit ($\tau_{\rm m}\ll 1$) was considered 
by Marscher (1983), 
who gave $b(\alpha)$ for $\alpha \geq -1.0$ 
(i.e., for $\tau_{\rm m}(0)<1.0$).

\begin{table*}
 \centering
 \begin{minipage}{140mm}
  \caption{\hspace{4pt}Table of constants}
  \begin{tabular}{@{}cccccccccc@{}}
  \hline
  \hline
  $\alpha$ & $0$
& $-$0.25 & $-$0.50
& $-$0.75 & $-$1.00
& $-$1.25 & $-$1.50
& $-$1.75 & $-$2.00 \\
  \hline
  $a$ & 0.2833
& 0.149 & 0.103
& 0.0831 & 0.0740
& 0.0711 & 0.0725
& 0.0776 & 0.0865 \\
  $C$ & 1.191
& 1.23 & 1.39
& 1.67 & 2.09
& 2.72 & 3.67
& 5.09 & 7.23 \\
  $\tau_{\rm m}(0)$
& 0
& 0.252 & 0.480
& 0.687 & 0.878
& 1.055 & 1.220
& 1.374 & 1.519 \\
  $\langle \tau_{\rm m} \rangle$
& 0
& 0.167 & 0.314
& 0.445 & 0.564
& 0.672 & 0.770
& 0.862 & 0.946 \\
  $b$ & 0
& 1.66 & 2.36
& 2.34 & 2.08
& 1.78 & 1.51
& 1.27 & 1.08 \\
  \hline
  \hline
\end{tabular}
\end{minipage}
\end{table*}


\subsection{Electron density}
\label{sec:lineC}

\begin{figure}
\centerline{ \epsfxsize=8cm \epsfbox[100 200 400 350]{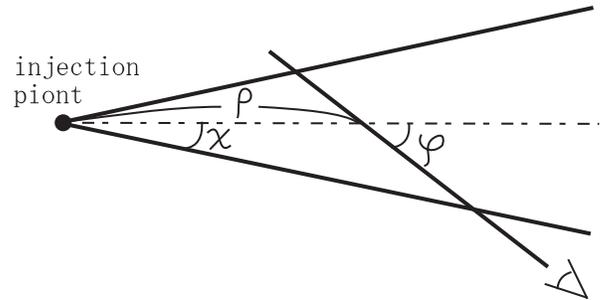} }
\caption{\label{fig:conical}
Schematic figure of a conical jet in the core region
with half opening angle $\chi$ in the observer's frame.
        }
\end{figure}

We next consider how to constrain $\Nco$
(i.e., line~C in fig.~\ref{fig:NeB})
from the theory of synchrotron self-absorption.
To examine the lower limit of $\Nco({\rm SSA})$, we consider
a conical jet geometry (fig.\ref{fig:conical}) in this section,
even though we apply the results
(eqs.~[\ref{eq:Nmin}]-[\ref{eq:def_e}])
to individual jet components, which would be more or less spherical 
rather than conical.
The optical depth $\tau$ for
synchrotron self absorption at distance $\rho$ from the injection point,
is given by
\begin{equation}
  \tau_{\nu}(\rho)
  = \left[ \frac{\rho \sin\chi}{\sin(\varphi +\chi)}
          +\frac{\rho \sin\chi}{\sin(\varphi -\chi)}
    \right]
    \alpha_{\nu},
\label{eq:tau1}
\end{equation}
where $\varphi$ is the viewing angle
and $\alpha_\nu$ [1/cm] refers to the effective absorption
(i.e., absorption minus stimulated emission) coefficient.
For a small half opening angle ($\chi \ll 1$),
this equation can be approximated as
\begin{equation}
  \tau_{\nu}(\rho)= 2 \frac{\rho \sin\chi}{\sin\varphi} \alpha_{\nu},
  \label{eq:tau2}
\end{equation}
where $\chi/\sin\varphi$ denotes the projected, observed
opening angle.
Noting that $\rho \sin\chi$ represents the transverse thickness
of the jet, 
we find that equation~(\ref{eq:tau2}) holds not only for 
a conical geometry but also for a cylindrical one.

Since $\tau_\nu$ and $\rho \sin\chi$ are Lorentz invariants, we obtain
\begin{equation}
  \frac{\alpha_\nu}{\sin\varphi}
  = \frac{\alpha_\nu^*}{\sin\varphi^*}.
  \label{eq:LI-1}
\end{equation}
Since $\nu \alpha_\nu$ is also Lorentz invariant,
equation (\ref{eq:LI-1}) gives
\begin{equation}
  \frac{\sin\varphi^*}{\sin\varphi}
  = \frac{\nu}{\nu^*}
  = \frac{\delta}{1+z} .
  \label{eq:LI-3}
\end{equation}
Combining equations (\ref{eq:tau2}) and (\ref{eq:LI-3}), we obtain
\begin{equation}
  \tau_\nu
      = \frac{1+z}{\delta} \frac{2\rho\sin\chi}{\sin\varphi} \alpha_\nu^*
      = \frac{1+z}{\delta} \frac{\theta_{\rm d}}{\sin\varphi}
        \frac{D_{\rm L}}{(1+z)^2} \alpha_\nu^* ,
  \label{eq:o-depth-2}
\end{equation}
where $D_{\rm L}/(1+z)^2$ is the angular diameter distance to the AGN;
$\theta_{\rm d}$ is measured in radian unit
(i.e., not in milliarcsecond).
Equation~(\ref{eq:o-depth-2}) holds 
for a conical geometry with an arbitrary small opening angle, 
including a cylindrical case.

We assume that the electron energy distribution
is represented by a power-law, 
\begin{equation}
  \frac{d\Nco}{d\gamma} = k_{\rm e}^\ast \gamma^{2\alpha-1}
  \qquad (\gamma_{\rm min} < \gamma < \gamma_{\rm max}),
  \label{eq:ele-dist}
\end{equation}
where $\Nco$ refers to the proper electron number density.
Integrating $d\Nco/d\gamma$ from $\gamma_{\rm min}$ to
$\gamma_{\rm max}$,
and assuming $\gamma_{\rm max} \gg \gamma_{\rm min}$
and $\alpha<0$, we obtain
\begin{equation}
  \Nco = \frac{\gamma_{\rm min}{}^{2\alpha}}{-2\alpha}
         k_{\rm e}^\ast.
  \label{eq:Ne-ke}
\end{equation}
The coefficient $C(\alpha)$ is given in Table~1 of Gould (1979)
and also in table~1 in this paper.

Substituting equation~(\ref{eq:abs-coeff}) into (\ref{eq:o-depth-2}),
and assuming $\gamma_{\rm min} \ll \gamma_{\rm max}$, we obtain
\begin{eqnarray}
  \Nco B^{-\alpha +1.5}
  &=& \frac{m_{\rm e}c}{e^2}
      \left( \frac{e}{2\pi m_{\rm e}c} \right)^{-1.5+\alpha}
      \frac{\tau_\nu(\alpha)}{C(\alpha)}
      \frac{\gamma_{\rm min}{}^{2\alpha}}{-2\alpha}
  \nonumber \\
  &&  \hspace{-1.5 truecm} \times
      \frac{\sin \varphi}{\theta_{\rm d}}
             \frac{(1+z)^2}{D_{\rm L}}
      \left( \frac{1+z}{\delta} \right)^{-\alpha+1.5}
      \nu^{-\alpha+2.5}.
  \label{eq:lineC-1}
\end{eqnarray}
Evaluating $\nu$ at $\nu_{\rm m}$,
and combining with equation (\ref{eq:lineB}),
we obtain $\Nco({\rm SSA})$ in cgs unit as
(see also Marscher 1983 for relatively optically thin cases, 
 $-\alpha\leq 1.0$)
\begin{eqnarray}
  \Nco({\rm SSA})
  &=& e(\alpha)
      \frac{\gamma_{\rm min}{}^{2\alpha}}{-2\alpha} 
      \frac{\sin\varphi}{d_{\rm A}}
      \left( \frac{\theta_{\rm d}}{\rm mas} \right)^{4\alpha-7}
   \nonumber \\
   & & \hspace*{-2.5 truecm} \times
      \left( \frac{\nu_{\rm m}}{\rm GHz}    \right)^{4\alpha-5}
      \left( \frac{S_{\rm m}}{\rm Jy}       \right)^{-2\alpha+3}
      \left( \frac{\delta}{1+z} \right)^{2\alpha-3},
   \label{eq:Nmin}
\end{eqnarray}
where
\begin{equation}
  d_{\rm A} 
  \equiv \frac{zq_0 +(q_0-1)(-1+\sqrt{2q_0 z+1})}{h (1+z)^2 q_0{}^2}
\end{equation}
\begin{equation}
  e(\alpha) \equiv
    1.71 \times 10^{-9}
    \times [ 2.79 \times 10^{-8} b(\alpha) ]^{\alpha-1.5}
    \times \frac{\tau_{\nu}}{C(\alpha)}
  \label{eq:def_e}
\end{equation}
If the jet component is discontinuous (e.g., spherical),
equation~(\ref{eq:Nmin}) gives the lower limit of 
$\Nco({\rm SSA})$,
because the path length becomes smaller than
$2\rho\sin\chi/\sin\varphi$ (eq.~[\ref{eq:tau2}]).

It looks like from equation~(\ref{eq:Nmin}) that 
$\Nco({\rm SSA})$
depends on $\theta_{\rm d}$ and $\nu_{\rm m}$ very strongly.
In the case of $\alpha=-0.75$, for instance, we obtain
$\Nco({\rm SSA}) \propto \theta_{\rm d}{}^{-10} \nu_{\rm m}{}^{-8}$.
Nevertheless, as a radio-emitting components evolves along the jet,
its $\theta_{\rm d}$ increases due to expansion
while its $\nu_{\rm m}$ decreases due to (synchrotron+adiabatic) cooling.
As a result, these two effects partially cancel each other and
suppress the variations of $\Nco({\rm SSA})$ along the jet.

\section{Kinetic luminosity in the core region}  
\label{sec:lineD}

In this section, we present a new method how to deduce the
kinetic luminosity, $L_{\rm kin}$, of a jet in the unresolved VLBI core
from the perpendicular core size,
utilizing the core-position offset among different frequencies 
as an intermediate variable in manipulation.
Once $L_{\rm kin}$ is obtained for the core,
we can apply it to derive lines~${\rm D}_1$ and ${\rm D}_2$
for individual jet components,
assuming a stationary jet ejection with constant $L_{\rm kin}$.

\subsection{Scaling Law}
\label{sec:scaling}

Let us introduce a dimensionless variable $r \equiv \rho/r_1$
to measure $\rho$ in parsec units, where $r_1=1$ pc.
We assume that the electron number density scales with $r$ as
\begin{eqnarray}
  \Nco = N_1 r^{-n},
  \label{eq:scaling_N}
\end{eqnarray}
where $N_1$ refers to the value of
$\Nco$ at $r=1$ (i.e., at 1 pc from the core in the observer's frame).
If particles are neither created nor annihilated in a conical jet,
conservation of particles gives $n=2$.
In the same manner, we assume the following scaling law:
\begin{eqnarray}
  B= B_1 r^{-m},
  \label{eq:scaling_B}
\end{eqnarray}
where $B_1$ refers to the values of $B$ at $r=1$.
In a stationary super-fast-magnetosonic jet, 
the magnetic flux conservation gives $m=1$. 
Note that we assume such scaling laws only in the unresolved VLBI core,
not in the jet region, for which we consider $\Nco({\rm SSA})$.

We further introduce the following dimensionless variables:
\begin{eqnarray}
  x_{\rm N} \equiv r_1 r_\circ{}^2 N_1,
  \nonumber \\
  x_{\rm B} \equiv \nu_{\rm B_1}/\nu_\circ
    = \frac{eB_1}{2\pi m_{\rm e}c} \cdot \frac{1}{\nu_0}.
  \label{eq:def_x}
\end{eqnarray}
Utilizing equation (\ref{eq:abs-coeff}),
and rewriting $k_{\rm e}^\ast$ and $\nu_{\rm B}$ in terms of
$x_{\rm N}$ and $x_{\rm B}$,
we obtain from the left equality in equation (\ref{eq:o-depth-2})
\begin{eqnarray}
  \tau_\nu
     &=& C(\alpha) \frac{2\chi}{\sin\varphi}
             \frac{-2\alpha}{\gamma_{\rm min}{}^{2\alpha}}
             \left(\frac{1+z}{\delta}\right)^{-\epsilon}
             \left(\frac{\nu}{\nu_\circ}\right)^{-1-\epsilon}
     \nonumber\\
     & & \times \quad
             r^{1-n-m\epsilon} x_{\rm N} x_{\rm B}{}^\epsilon,
  \label{eq:tau4}
\end{eqnarray}
where $\epsilon \equiv 3/2 -\alpha$.
Note that both $x_{\rm N}$ and $x_{\rm B}$ should be distinguished
from $\Nco({\rm SSA})$ and $B$ presented in section~\ref{sec:SSA},
because we are considering the core region in this section.

At a given frequency $\nu$,
the flux density will peak at the position where $\tau_\nu$ becomes unity.
Thus setting $\tau=1$ and solving equation (\ref{eq:tau4})
for $r$, we obtain the distance from the VLBI core
observed at frequency $\nu$ from the injection point as (Lobanov 1998)
\begin{equation}
  r(\nu) = \left[ x_{\rm B}{}^{k_{\rm b}} f(\alpha) \frac{\nu_\circ}{\nu}
           \right]^{1/k_{\rm r}}
  \label{eq:core_rad}
\end{equation}
where
\begin{equation}
  f(\alpha) \equiv \left[ C(\alpha) \frac{2\chi}{\sin\varphi}
                          \frac{-2\alpha}{\gamma_{\rm min}{}^{2\alpha}}
                          \left(\frac{\delta}{1+z} \right)^{\epsilon}
                          x_{\rm N}
                   \right]^{1/(\epsilon+1)}
\end{equation}
\begin{equation}
  k_{\rm b} \equiv \frac{3-2\alpha}{5-2\alpha},
\end{equation}
\begin{equation}
  k_{\rm r} \equiv \frac{(3-2\alpha)m+2n-2}{5-2\alpha}.
\end{equation}

\subsection{Core-Position Offset}
\label{sec:CPO}

If we measure $r(\nu)$ at  two different frequencies
(say $\nu_{\rm a}$ and $\nu_{\rm b}$),
equation (\ref{eq:core_rad}) gives the dimensionless, projected
distance of
$r(\nu_{\rm a})-r(\nu_{\rm b})$ as
\begin{eqnarray}
  \Delta r_{\rm proj}
  &=& \left[ r(\nu_{\rm a}) - r(\nu_{\rm b}) \right] \sin\varphi
  \nonumber\\
  &=& (x_{\rm B}{}^{k_{\rm b}} f \nu_\circ)^{1/k_{\rm r}}
      \frac{\nu_{\rm b}^{1/k_{\rm r}} - \nu_{\rm a}^{1/k_{\rm r}}}
      {\nu_{\rm a}^{1/k_{\rm r}} \nu_{\rm b}^{1/k_{\rm r}}}
      \sin\varphi.
  \label{eq:del_jet}
\end{eqnarray}
Here, $\Delta r_{\rm proj}$ is in pc units;
therefore, $r_1 \Delta r_{\rm proj}$ represents the projected
distance of the two VLBI cores (in cm, say).
That is, $r_1 \Delta r_{\rm proj}$ equals
$4.85 \cdot 10^{-9} \Delta r_{\rm mas} D_{\rm L}/(1+z)^2$
in equation~(4) in Lobanov (1998),
where $\Delta r_{\rm mas}$ refers to the
core-position difference in mas.
Defining the core-position offset as 
\begin{equation}
  \Omega_{r \nu} \equiv
    r_1 \Delta r_{\rm proj}
    \frac{\nu_{\rm a}^{1/k_{\rm r}} \nu_{\rm b}^{1/k_{\rm r}}}
         {\nu_{\rm b}^{1/k_{\rm r}} - \nu_{\rm a}^{1/k_{\rm r}}},
  \label{eq:def_CPO}
\end{equation}
we obtain
\begin{equation}
  \frac{\Omega_{r\nu}}{r_1}
  = (x_{\rm B}^{k_{\rm b}} f \nu_\circ )^{1/k_{\rm r}} \sin\varphi
  \label{eq:CPO}
\end{equation}
Setting $\nu_{\rm b} \rightarrow \infty$ in equation (\ref{eq:del_jet}),
we can express the absolute distance of the VLBI core
measured at $\nu$ from the central engine as (Lobanov 1998)
\begin{equation}
  r_{\rm core} (\nu) = \frac{\Omega_{r \nu}}{r_1 \sin \varphi}
                       \nu^{-1/k_{\rm r}}.
  \label{eq:Rcore}
\end{equation}
To express $x_{\rm B}$ in terms of $x_{\rm N}$ and
$\Omega_{r\nu}$, we solve equation (\ref{eq:CPO})
for $x_{\rm B}$ to obtain
\begin{equation}
  x_{\rm B} = \left( \frac{\Omega_{r\nu}}{r_1 \sin\varphi}
              \right)^{k_{\rm r}/k_{\rm b}}
              (f \nu_\circ)^{-1/k_{\rm b}}.
  \label{eq:xB}
\end{equation}
Note that $x_{\rm N}$ is included in $f=f(\alpha)$.

We next represent $x_{\rm N}$ and $x_{\rm B}$
(or equivalently, $N_1$ and $B_1$) as a function of $\Omega_{r\nu}$.
To this end, we parameterize the energy ratio
between particles and the magnetic field as
\begin{equation}
  \Nco \gamma_{\min} m_{\rm e} c^2 = K \frac{B^2}{8 \pi}.
  \label{eq:equiP_0}
\end{equation}
When an energy equipartition between the radiating particles
and the magnetic field holds,
we obtain
\begin{equation}
  K = \frac{\gamma_{\rm min}}{\langle \gamma_- \rangle},
  \label{eq:K_1}
\end{equation}
where the averaged electron Lorentz factor, $\langle \gamma_- \rangle$
becomes
\begin{eqnarray}
  \langle \gamma_- \rangle
    &\equiv&
        \frac{ \displaystyle{\int_{\gamma_{\rm min}}^{\gamma_{\rm max}}
               \gamma \cdot k_{\rm e}^\ast \gamma^{2\alpha-1} d\gamma }}
             { \displaystyle{\int_{\gamma_{\rm min}}^{\gamma_{\rm max}}
                            k_{\rm e}^\ast \gamma^{2\alpha-1} d\gamma }}
    \nonumber \\
    &=& \frac{2\alpha}{2\alpha+1} \gamma_{\rm min}
        \frac{(\gamma_{\rm max}/\gamma_{\rm min})^{2\alpha+1}-1}
             {(\gamma_{\rm max}/\gamma_{\rm min})^{2\alpha  }-1}.
  \label{eq:def_gamAVR}
\end{eqnarray}
for $\alpha<0$.
We assume a rough energy equipartition holds and adopt
\begin{equation}
  K \sim \frac{2\alpha+1}{2\alpha}
         \frac{(\gamma_{\rm max}/\gamma_{\rm min})^{2\alpha  }-1}
              {(\gamma_{\rm max}/\gamma_{\rm min})^{2\alpha+1}-1}.
\end{equation}
In the limit $\alpha \rightarrow -0.5$, the right-hand side tends to
$ 1 / \ln(\gamma_{\rm max} / \gamma_{\rm min}) $,
which is $0.1$ if $\gamma_{\rm max}=10^{4.34}\gamma_{\rm min}$,
for instance.
Note that we assume a rough energy equipartition in the core
and not in the jet.

In this paper, we assume that $K/\gamma_{\rm min}$ is constant 
for $r$.
Then, equation~(\ref{eq:equiP_0}) requires $n=2m$.
Substituting $N_{\rm e}^* = N_1 r^{-2m}$ and $B = B_1 r^{-m}$ into
equation~(\ref{eq:equiP_0}),
and replacing $N_1$ and $B_1$ with $x_{\rm N}$ and $x_{\rm B}$,
we obtain
\begin{equation}
  x_{\rm N}= \frac{\pi}{2} \frac{K}{\gamma_{\rm min}}
             \frac{r_1}{r_\circ}
             x_{\rm B}{}^2
  \label{eq:equiP_1}
\end{equation}
Combining equations (\ref{eq:xB}) and (\ref{eq:equiP_1}),
we obtain
\begin{eqnarray}
  x_{\rm B}
  &=& \left[ \frac{1}{\nu_\circ}
             \left(\frac{\Omega_{r\nu}}{r_1 \sin\varphi}\right)^{k_r}
      \right]^{(5-2\alpha)/(7-2\alpha)}
  \nonumber \\
  && \hspace{-1.0 truecm} \times
     \left[ \pi C(\alpha) \frac{\chi}{\sin\varphi}
             \frac{K}{\gamma_{\rm min}}
             \frac{r_1}{r_\circ}
             \frac{-2\alpha}{\gamma_{\rm min}{}^{2\alpha}}
             \left(\frac{\delta}{1+z}\right)^\epsilon
      \right]^{-2/(7-2\alpha)}.
  \label{eq:sol_xB}
\end{eqnarray}
For ordinary values of $\alpha(<0)$,
$x_{\rm B}$ decreases with increasing $K$, as expected.
The particle number density, $x_{\rm N}$, can be
readily computed from equation (\ref{eq:equiP_1}).

\subsection{Kinetic Luminosity}          
\label{sec:Lkin}

When the jet has a cross section $\pi R_\perp^2$ at a
certain position,
$L_{\rm kin}$ and $\Nco$ are related by
\begin{equation}
   L_{\rm kin}
   =  \pi R_\perp{}^2 \beta c \cdot \Gamma \Nco \cdot (\Gamma-1)
      \left( \langle\gamma_-\rangle m_{\rm e}c^2
            +\langle\gamma_+\rangle m_+ c^2
      \right) ,
   \label{eq:Lkin}
\end{equation}
where $\langle\gamma_+\rangle$ refers to the
averaged Lorentz factors of positively charged particles,
and $m_+$ designates the mass of the positive charge.

Assuming that the particle number is conserved, we find that
$\pi R_\perp{}^2 \beta c \cdot \Gamma \Nco$ 
is constant along a stationary jet.
Thus, provided that $\Gamma \gg 1$ holds, 
$R_\perp{}^2 \Nco$ can be replaced as
\begin{equation}
  R_\perp{}^2 \Nco 
    = \frac{\Gamma_1}{\Gamma} (r_1 \chi_1)^2 N_1 
    = \frac{\Gamma_1}{\Gamma} \chi_1{}^2 \frac{r_1}{r_\circ{}^2} x_{\rm N},
  \label{eq:NR2}
\end{equation}
where $\Gamma_1$ and $\chi_1$ refer to $\Gamma$ and $\chi$ at 1~pc.
Substituting equation (\ref{eq:NR2}) into (\ref{eq:Lkin}),
we obtain 
\begin{eqnarray}
  L_{\rm kin} 
  &=& C_{\rm kin} K \frac{r_1{}^2}{r_\circ{}^3} 
      \beta \Gamma_1 (\Gamma-1) 
      \left(\frac{\chi_1}{\chi}\right)^2 
      \nonumber \\
  & & \hspace{-1.5 truecm} \times
      \left[ \frac{\chi}{\nu_\circ} 
             \left(\frac{\Omega_{r\nu}}{r_1 \sin\varphi}\right)^{k_r}
      \right]^{2(5-2\alpha)/(7-2\alpha)}
      \nonumber \\ 
  & & \hspace{-1.5 truecm} \times 
      \left[ \frac{\pi C(\alpha)}{\sin\varphi}
             \frac{K}{\gamma_{\rm min}} \frac{r_1}{r_\circ}
             \frac{-2\alpha}{\gamma_{\rm min}{}^{2\alpha}}
             \left(\frac{\delta}{1+z}\right)^{3/2-\alpha}
      \right]^{-4/(7-2\alpha)} \hspace*{-1.5truecm};\hspace*{1.0truecm}
  \label{eq:Lkin_CPO}
\end{eqnarray}
we adopt
$C_{\rm kin}= \pi^2 \langle\gamma_-\rangle m_{\rm e}c^3/\gamma_{\rm min}$
for a pair plasma, while
$C_{\rm kin}= \pi^2 \langle\gamma_+\rangle m_{\rm p}c^3/(2\gamma_{\rm min})$
for a normal plasma,
where $m_{\rm p}$ refers to the rest mass of a proton.
Equation~(\ref{eq:Lkin_CPO}) holds
not only for a conical jet but also for a general jet with $\Gamma \gg 1$
provided that the particle number is conserved
and that $n=2m$ holds.
It should be noted that $\gamma_{\rm min}$ takes different values
between a pair and a normal plasma.

\subsection{Jet Opening Angle}      
\label{sec:opening}

Let us now consider the half opening angle of the jet,
and further reduce the expression of $L_{\rm kin}$.
As noted in \S~\ref{sec:Lkin},
we obtain
\begin{equation}
  \chi = \frac{R_\perp}{\rho}
       = \frac{0.5\theta_{\rm d,core}(\nu)}{\rho}
         \frac{D_{\rm L}}{(1+z)^2}
  \label{eq:chi}
\end{equation}
for a conical geometry,
where $\theta_{\rm d,core}(\nu)$ is the angular diameter of the VLBI core
at a frequency $\nu$ in the perpendicular direction to the
jet-propagation direction on the projected plane (i.e., a VLBI map).
The luminosity distance is given by
\begin{equation}
  D_{\rm L}= 4.61 \times 10^9 h^{-1} r_1
             \frac{q_0 z +(q_0-1)(-1+\sqrt{2q_0 z+1})}{q_0{}^2}.
  \label{eq:Da}
\end{equation}
On the other hand, equation (\ref{eq:Rcore}) gives
\begin{equation}
  \rho
    = r_{\rm core} r_1
    = \frac{\Omega_{r \nu}}{\sin\varphi} \nu^{-1/k_{\rm r}}
  \label{eq:Rcore_2}
\end{equation}
Substituting equations~(\ref{eq:Rcore_2}), (\ref{eq:Da}) into
(\ref{eq:chi}), we obtain
\begin{eqnarray}
  \frac{1}{\nu_0} \frac{\chi\Omega_{r\nu}}{r_1 \sin\varphi}
  &=& 11.1 h^{-1} \frac{q_0 z +(q_0-1)(-1+\sqrt{2q_0 z+1})}
                        {q_0{}^2 (1+z)^2}
  \nonumber\\
  & & \times
    \left( \frac{\theta_{\rm d,core}}{\rm mas} \right)
    \frac{\nu^{1/k_r}}{\nu_0}.
  \label{eq:chi_CPO}
\end{eqnarray}

Since $m=1$ is consistent with the theory of magnetohydrodynamics,
we will adopt $k_r=1$ in the rest of this paper.
Then, it follows that the factor containing 
$\chi\Omega_{r\nu}/\sin\varphi$
in equation (\ref{eq:Lkin_CPO}) can be simply expressed
in terms of $\theta_{\rm d,core}(\nu)$.
That is, to evaluate $L_{\rm kin}$,
we do not have to stop on the way at the computation of
$\Omega_{r\nu}$,
which requires more than two VLBI frequencies.
We only need the VLBI-core size at a single frequency,
$\theta_{\rm d,core}(\nu)$,
in addition to $\Gamma$, $\varphi$, $\alpha$, $\gamma_{\rm min}$,
$z$, $K$, and $C_{\rm kin}$.
If we have to examine $\chi$,
we additionally need $\Omega_{r\nu}$.
The information of the composition is included in $C_{\rm kin}$.
In the present paper, we assume $\Gamma\chi^2=\Gamma_1 \chi_1{}^2$.

If $k_r$ deviates from unity, we have to independently examine
$\chi$ and $\Omega_{r\nu}$. 
In another word, we need at least two VLBI frequencies
(e.g., $2$ and $8$ GHz) to measure the core-position offset
in order to constrain $L_{\rm kin}$, 
even if we know the value of $k_r$ by some means.

To constrain $\Gamma$ in equation~(\ref{eq:Lkin_CPO}), 
let us briefly consider how to obtain its lower bound.
If the apparent velocity of a radio-emitting component, 
$\beta_{\rm app}=\beta\sin\varphi/(1-\beta\cos\varphi)$,
reflects the fluid velocity,
the lower bound can be obtained by
$\Gamma_{\rm min} \equiv \sqrt{\beta_{\rm app}{}^2+1}$.
On the other hand, the upper bound of $\Gamma$
is not easily constrained; 
thus, we must consider it source by source,
using observations at different photon energies (e.g., X-rays)
if necessary.

\subsection{Electron density deduced from kinetic luminosity}
\label{sec:Ne_kin}

Once $L_{\rm kin}$ is known by the method described above
or by some other methods,
we can compute the electron number density as a function of the
composition.
We assume that the $L_{\rm kin}$ deduced for the core
can be applied for the individual jet components.
Then, solving equation (\ref{eq:Lkin}),
which holds not only in the core but also in the jet,
for $\Nco$,
and utilizing $R_\perp = (\theta_{\rm d}/2)D_{\rm L}/(1+z)^2$,
we obtain for a normal plasma
\begin{eqnarray}
  \Nco({\rm nml})
  &=& 5.93 \times 10^{-2} h^2 \frac{L_{46}}{\Gamma(\Gamma-1)}
      \left( \frac{\theta_{\rm d}}{\rm mas} \right)^{-2}
  \nonumber \\
  & & \times \left[ \frac{q_0{}^2 (1+z)^2}
                     {zq_0+(q_0-1)(-1+\sqrt{2q_0z+1})}
         \right]^2
  \nonumber \\
  & & \times \frac{1}{\langle\gamma_+\rangle}
         \frac{1}{1+\langle\gamma_-\rangle m_{\rm e}/m_{\rm p}}
         \, \mbox{cm}^{-3},
  \label{eq:Ne_nml}
\end{eqnarray}
where $L_{46}= L_{\rm kin}/(10^{46} \mbox{ergs s}^{-1})$.
If $\Nco({\rm nml})$ becomes less than the electron density deduced
independently from the theory of synchrotron self-absorption (SSA),
the possibility of a normal plasma dominance can be ruled out.

It is worth comparing $\Nco({\rm nml})$ with $\Nco({\rm pair})$,
the particle density in a pure pair plasma.
If we compute $L_{\rm kin}$ by using equation~(\ref{eq:Lkin_CPO})
and (\ref{eq:chi_CPO}), we obtain the following ratio
\begin{eqnarray}
  \frac{\Nco({\rm pair})}{\Nco({\rm nml})}
  &=& \frac{L_{\rm kin}({\rm pair})}{L_{\rm kin}({\rm nml})}
      \frac{m_{\rm p}c^2
            +\langle\gamma_-({\rm nml })\rangle m_{\rm e}c^2}
           {2\langle\gamma_-({\rm pair})\rangle m_{\rm e}c^2}
  \nonumber \\
  &=& \frac{\gamma_{\rm min}({\rm nml })}
           {\gamma_{\rm min}({\rm pair})},
  \label{eq:DensRat}
\end{eqnarray}
where $\gamma_{\rm min}({\rm pair})$ and $\gamma_{\rm min}({\rm nml})$
refer to $\gamma_{\rm min}$ in a pair and a normal plasma, respectively;
$\langle \gamma_-({\rm pair}) \rangle$ and
$\langle \gamma_-({\rm nml})  \rangle$
refer to the averaged Lorentz factor of electrons in a
pair and a normal plasma, respectively.
Therefore, if
$\gamma_{\rm min}({\rm pair})$ is comparable to
$\gamma_{\rm min}({\rm nml}) \sim 100$,
there is little difference between
$\Nco({\rm pair})$ and $\Nco({\rm nml})$.



\subsection{Summary of the Method}

Let us summarize the main points that have been made 
in \S\S~\ref{sec:SSA} and \ref{sec:lineD}. \\
(1) \
We can compute the magnetic field strength 
(i.e., line~B in fig.~\ref{fig:NeB})
from the surface brightness condition.\\
(2) \
We can also constrain the proper electron density and the
magnetic field strength (i.e., line~C)
from the synchrotron self-absorption constraint.\\
(3) \
Combing (1) and (2), we obtain the electron density,
$\Nco({\rm SSA})$, for each radio-emitting component.\\
(4) \ 
Observing the core size, $\theta_{\rm d,core}(\nu)$, 
at a single frequency, $\nu$,
and substituting equation~(\ref{eq:chi_CPO}) into 
equation~(\ref{eq:Lkin_CPO}),
we obtain the kinetic luminosity, $L_{\rm kin}$.\\
(5) \
Assuming a normal-plasma dominance,
we obtain the proper electron density, $\Nco({\rm nml})$
(i.e., line~${\rm D}_2$) 
from $L_{\rm kin}$ by equation~(\ref{eq:Ne_nml}).\\
(6) \
If $\Nco({\rm nml}) \ll \Nco({\rm SSA})$ holds,
we can rule out the possibility of a normal-plasma dominance and
that of a pair-plasma dominance with $\gamma_{\rm min} \sim 100$ or greater.
That is, $\Nco({\rm nml}) \ll \Nco({\rm SSA})$ indicates the
dominance of a pair plasma with $\gamma_{\rm min} \ll 100$.


\section{Application to the 3C~345 Jet}    
\label{sec:3C345_appli}

Let us apply the method described in the previous sections
to the radio-emitting components in the 3C~345 jet
and investigate the composition.
This quasar ($z=0.595$; Hewitt \& Burbidge 1993)
is one of the best studied objects showing structural and spectral
variabilities on parsec scales around the compact unresolved core
(for a review, see e.g., Zensus 1997).
At this redshift, 1 mas corresponds to $5.85 h^{-1}$ pc,
and 1 mas ${\rm yr}^{-1}$ to $\beta_{\rm app}= 30.3 h^{-1}$.

\subsection{Kinetic luminosity}
\label{sec:3C345_Lkin}

\begin{figure}
\centerline{ \epsfxsize=9cm \epsfbox[200 20 500 250]{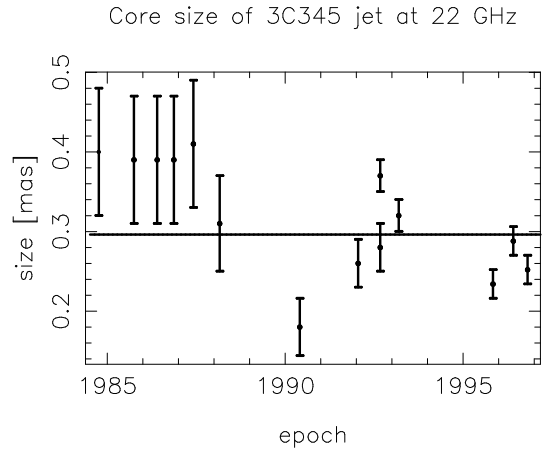} }
\caption{\label{fig:core22}
Angular core size of the 3C~345 core at 22.2 GHz.
The horizontal line represents the bet fit value.
        }
\end{figure}

To deduce $L_{\rm kin}$, we first consider
$\theta_{\rm d, core} \nu / \nu_0$,
where $k_r=1$ is adopted in equation~(\ref{eq:chi_CPO}).
From the reported core size at 22.2 GHz
    at 6 epochs by Zensus et al. (1995),
    at 5 epochs by Unwin  et al. (1997),
and at 3 epochs by Ros    et al. (2000),
we can deduce the averaged core size, $\theta_{\rm d,core}$,
as $0.296$ mas (fig.~\ref{fig:core22});
here, we evaluate the core size with $1.8\sqrt{ab}$ for
the latter three epochs,
where $a$ and $b$ refer to the major and minor axes at the FWHM
of the elliptical Gaussian presented in Ros et al. (2000).
From $\theta_{\rm d,core}= 0.296$ mas, we obtain
$\theta_{\rm d,core} \nu / \nu_0 = 6.18 \times 10^{-14}$
as the averaged value over the 14 epochs at 22 GHz.
In the same manner, we obtain the following averaged values
at different frequencies:
$\theta_{\rm d,core}= 0.258$ mas and
$\theta_{\rm d,core} \nu / \nu_0 =  3.64 \times 10^{-14}$
over 3 epochs at 15 GHz (Ros et al. 2000);
$\theta_{\rm d,core}= 0.436$ mas and
$\theta_{\rm d,core} \nu / \nu_0 =  4.38 \times 10^{-14}$
over 7 epochs at 10.7 GHz
(1 epoch from Unwin et al. 1994;
 4 epochs from Zensus et al. 1995;
 2 epochs from Gabuzda et al. 1999);
$\theta_{\rm d,core}= 0.343$ mas and
$\theta_{\rm d,core} \nu / \nu_0 =  2.71 \times 10^{-14}$
over 7 epochs at 8.4 GHz
(1 epoch from Unwin et al. 1994;
 3 epochs from Zensus et al. 1995;
 3 epochs from Ros et al. 2000);
$\theta_{\rm d,core}= 0.423$ mas and
$\theta_{\rm d,core} \nu / \nu_0 =  1.98 \times 10^{-14}$
over 11 epochs at 5.0 GHz
(4 epochs from Brown et al. 1994;
 1 epoch from Unwin et al. 1994;
 3 epochs from Zensus et al. 1995;
 3 epochs from Ros et al. 2000).
Taking a weighted average of $\theta_{\rm d,core} \nu / \nu_0$
over the 42 epochs at the different 5 frequencies, we obtain
\begin{equation}
  \frac{\theta_{\rm d,core} \nu}{\nu_0}= 4.01 \times 10^{-14}.
  \label{eq:core_1}
\end{equation}

We next consider the spectral index
that reflects the energy distribution of the power-law electrons
in the core. 
From infrared to optical observations
(J,H,K,L bands from Sitko et al. 1982;
 J,H,K   bands from Neugebauer et al. 1982;
 K       band  from Impey et al. 1982;
 IRAS 12, 25, 60, 100 microns from Moshir et al. 1990),
we obtain
\begin{equation}
  \alpha= -1.15
  \label{eq:core_2}
\end{equation}
for the core.
This value is consistent with that adopted by Zensus et al. (1995).

Using equations~(\ref{eq:core_1}) and (\ref{eq:core_2}), we obtain
from equation~(\ref{eq:chi_CPO})
\begin{equation}
  \left( \frac{1}{\nu_0}
         \frac{\chi \Omega_{r\nu}}{r_1 \sin\varphi}
  \right)^{2(5-2\alpha)/(7-2\alpha)}
  = 4.97 \times 10^{-21} h^{-1.57}.
  \label{eq:core_3}
\end{equation}
Moreover, $\alpha=-1.15$ gives
\begin{eqnarray}
  \, & &
  \left[ \pi C(\alpha) \frac{-2\alpha K}{\gamma_{\rm min}{}^{2\alpha+1}}
         \frac{r_1}{r_0}
  \right]^{-4/(7-2\alpha)}
  \nonumber \\
  &=& 1.31 \times 10^{-14} K^{-0.43} \gamma_{\rm min}{}^{-0.56}.
  \label{eq:core_4}
\end{eqnarray}
We thus obtain
\begin{eqnarray}
  \lefteqn{
  L_{\rm kin}
  = 4.68 \times 10^{45} h^{-1.57}
      \langle\gamma_+\rangle
      \left(\frac{\gamma_{\rm min}}{100}\right)^{-1.56} K^{0.57}}
  \nonumber\\
  & & \hspace*{-1.0 truecm}
  \times \Gamma(\Gamma-1)
             \left[ \frac{1}{\sin\varphi}
                    \left(\frac{\delta}{1+z}\right)^{2.65}
             \right]^{-0.43}
      \mbox{ergs s}^{-1}
  \label{eq:Lkin_3C345_0}
\end{eqnarray}
for a normal plasma, that is,
$C_{\rm kin}=\pi^2\langle\gamma_+\rangle m_{\rm p}c^3/(2\gamma_{\rm min})$
in equation~(\ref{eq:Lkin_CPO}).
Note that the factor $\langle\gamma_+\rangle$
cancels out when we compute $\Nco({\rm nml})$ 
by equation~(\ref{eq:Ne_nml})
and hence does not affect the conclusions on the matter content
in the present paper.
However, to examine the absolute kinetic luminosity
of a normal-plasma-dominated jet,
we have to independently know $\langle\gamma_+\rangle$,
because we can constrain only the electron density by SSA theory.

\subsection{Kinematics near the core}
\label{sec:3C345_kinematics}

To constrain $L_{\rm kin}$ further,
we must consider the kinematics of the core
and evaluate $\Gamma$ and $\varphi$.
In the case of the 3C~345 jet,
components are found to be \lq accelerated' as they propagate
away from the core.
For example, for component C4,
$\beta_{\rm app}h<8$ held when the distance from the core was
less than 1 mas (Zensus et al. 1995);
however, it looked to be \lq accelerated' from this distance
(or epoch 1986.0)
and attained $\beta_{\rm app}h \sim 45$
at epoch 1992.5 (Lobanov \& Zensus 1994).
It is, however, unlikely that the fluid bulk Lorentz factor increases
so much during the propagation.
If $\beta_{\rm app}$ reflected the fluid bulk motion of C4,
the bulk Lorentz factor, $\Gamma$, had to be at least $45$
until 1992.5.
Then $\beta_{\rm app} \sim 6$ during 1981-1985 indicates
that the viewing angle should be less than $0.05^\circ$,
which is probably much less than the opening angle.
Analogous \lq acceleration' was reported also for C3 (Zensus et al. 1995)
and C7 (Unwin et al. 1997).

On these grounds,
we consider the apparent \lq acceleration' of the components
in their later stage of propagation on VLBI scales
are due to geometrical effects such as \lq scissors effects'
(Hardee \& Norman 1989; Fraix-Burnet 1990),
which might be created at the intersection of a pair of shock waves.
Such shock waves may be produced, for example,
by the interaction between relativistic pair-plasma beam and
the ambient, non-relativistic normal-plasma wind
(Sol, Pelletier \& Ass$\acute{\rm e}$o 1989;
 Pelletier \& Roland 1989; Pelletier \& Sol 1992;
 Despringre \& Fraix-Burnet 1997;
 Pelletier \& Marcowith 1998). 
In their two-fluid model,
the pair beam could be destructed
at a certain distance from the core on VLBI scales
due to the generation of Langmuir waves.

Then, how should we constrain $\Gamma$ and $\varphi$ for the 3C~345 jet?
In this paper, we assume that the radio-emitting components
before their rapid acceleration represent
(or mimic) the fluid bulk motion.
Under this assumption,
Steffen et al. (1995) fitted the motions of C4
(during 1980-1986) and C5 (during 1983-1989)
when the components are within 2 mas from the core by a helical model.
They found $\Gamma= 5.8$ for C4 and $4.6$ for C5
with $\varphi= 6.8^\circ$, assuming $h=1.54$ (i.e., $H_0= 100$km/s/Mpc).
These Lorentz factors will be $\sim 9$ and $\sim 7$ for C4 and C5,
respectively,
if $h=1$ (i.e., $H_0=65$km/s/Mpc).
Subsequently, Qian et al. (1996) investigated the intrinsic evolution
of C4 under the kinematics of $\Gamma=5.6$ and
$\varphi=2^\circ$ -- $8^\circ$, assuming h= 1.54;
the range of viewing angles is in good agreement with
$\varphi=2^\circ$ (for C4) -- $4^\circ$ (for C2) at 1982.0
obtained by Zensus et al. (1995),
who used a larger value of $\Gamma=10$ under $h=1.54$,
which corresponds to $\Gamma \sim 15$ for $h=1$.

Unless a substantial deceleration takes place, the assumption of a
constant Lorentz factor is justified as the first order of approximation.
In the last part of \S~\ref{sec:3C345_individual}, 
we will examine the case when the
Lorentz factor is different from this assumed value
and consider how the conclusion of the composition depends on $\Gamma$.
Close to the core, we adopt $\varphi=2^\circ$,
because the viewing angle is suggested to decrease with decreasing
distance from the core (Zensus et al. 1995; Unwin et al. 1997).
If $\varphi$ is less than $2^\circ$ in the core,
$L_{\rm kin}$, and hence $\Nco({\rm nml})$ further decreases;
that is, the normal-plasma dominance can be further ruled out.
In short, we apply $K=0.1$, 
$\Gamma=15$ and $\varphi=2^\circ$ in equation~(\ref{eq:Lkin_3C345_0}).
Assuming a normal-plasma dominance, we adopt
$\gamma_{\rm min}=10^2$ for electron energy distribution. 

\subsection{Composition of individual components}
\label{sec:3C345_individual}

We can now investigate the composition of the radio-emitting
components in the 3C~345 jet, using their spectral information.
We assume that the jet is neither accelerated nor decelerated and
apply the Lorentz factor obtained for the (unresolved) core
to all the (resolved) jet components.
To compute $\Nco({\rm SSA})$, we further need $\varphi$
for each component at each epoch.
(Note that $\varphi=2^\circ$ is assumed for the core,
not for the components.)
For this purpose, we assume that $\beta_{\rm app}$ reflects
the fluid motion if $\beta_{\rm app}<\sqrt{\Gamma^2-1}$ and
that $\varphi$ is fixed after $\beta_{\rm app}$
exceeds $\sqrt{\Gamma^2-1}$.
That is, we compute $\varphi$ from
\begin{eqnarray}
  \tan\varphi
  = \left\{ \begin{array}{ll}
              \frac{\displaystyle 2\beta_{\rm app}}
                   {\displaystyle \beta_{\rm app}{}^2+\delta^2-1}
                & \mbox{if $\beta_{\rm app}<\sqrt{\Gamma^2-1}$} \\
              \ & \\
              \frac{\displaystyle 1}
                   {\displaystyle \sqrt{\Gamma^2-1}}
                & \mbox{if $\beta_{\rm app}>\sqrt{\Gamma^2-1}$}
            \end{array}
    \right.
    \label{eq:view_ang}
\end{eqnarray}
The constancy of $\varphi$
when $\beta_{\rm app}>\sqrt{\Gamma^2-1}=15.0$ could be justified
by the helical model of Steffen et al. (1995),
who revealed that the viewing angle does not vary significantly
after the component propagates a certain distance
(about 1 mas in the cases of components C4 and C5) from the core.
As we will see in table~2 (later in this subsection),
$\varphi$ increases with increasing distance from the core
and do not deviate significantly from the
assumed value for the core ($2$ degree)."

The synchrotron self-absorption spectra of individual components are
reported in several papers.
They were first reported in Unwin et al. (1994),
who presented ($\nu_{\rm m}$,$S_{\rm m}$,$\alpha$) and $\theta_{\rm d}$
(or $\xi$ in their notation) of components C4 and C5 at epoch 1990.7.
Subsequently, Zensus et al. (1995) gave
($\nu_{\rm m}$,$S_{\rm m}$,$\alpha$), $\theta_{\rm d}$,
and $\beta_{\rm app}h$
for C2, C3, and C4 at 1982.0.
However, the errors are given only for $\theta_{\rm d}$; therefore,
we evaluate the errors of the output parameters,
$\delta$, $\varphi$, $B$, $\Nco({\rm nml})$,
and $\Nco({\rm nml})/\Nco({\rm SSA})$,
using the errors in $\theta_{\rm d}$ alone for C2, C3, and C4 at 1982.0.
Later, Unwin et al. (1997) presented
($\nu_{\rm m}$,$S_{\rm m}$,$\alpha$), $\theta_{\rm d}$,
and $\beta_{\rm app}h$
of C5 (at 1990.55) and C7 (at 1992.05, 1992.67, 1993.19, and 1993.55).
More recently, Lobanov and Zensus (1999) presented a comprehensive data
of ($\nu_{\rm m}$,$S_{\rm m}$,$\alpha$) of C3, C4, and C5 at various
epochs.
We utilize $\theta_{\rm d}= 0.114 + 0.0658({\rm epoch}-1979.50)$ for C4
(Qian et al. 1996),
$\theta_{\rm d}= 0.114 + 0.0645({\rm epoch}-1980.00)$ for C5,
which is obtained from the data given in Zensus et al. (1995) and
Unwin et al. (1997),
and $\theta_{\rm d}= 0.09 + 0.45(\rho\sin\varphi/{\rm mas})$ for C3
(Biretta et al. 1986),
where $\rho\sin\varphi$ refers to the projected distance from the core.

The results for individual components at various epochs are given
in table~2 for $K=0.1$.
It follows that the ratio, $\Nco({\rm nml}) / \Nco{}({\rm SSA})$
becomes less than $1$ if the magnetic field is an acceptable strength,
(e.g., $B<100{\rm mG}$).
For C4, we cannot rule out the possibility of a normal-plasma dominance
at epochs 1985.8 and 1988.2.
Nevertheless, at these two epochs,
unnatural input parameters
($\alpha=-0.2$ at epoch 1985.8 and  $S_{\rm m}=0.8$~Jy at epoch 1988.2)
give too large $B$ $(>380 \mbox{mG})$ for a moderate $\delta (<30)$.
If we instead interpolated $S_{\rm m}$ and $\alpha$ at epochs
1985.8 and 1988.2 from those at epochs 1983.4 and 1990.7,
we would obtain reasonable values
$B=70$~mG and $\Nco({\rm nml})/\Nco({\rm SSA})=0.28$
at epoch 1985.8, and
$B=50$~mG and $\Nco({\rm nml})/\Nco({\rm SSA})=0.046$
at epoch 1988.2.
On the other hand, C5 at 1984.2 gives unusual value
$\Nco({\rm SSA})=3.7\times 10^5 \mbox{cm}^{-3}$.
It is due to the unnaturally small value of $\nu_{\rm m}$($=2.5$~GHz)
in the early stage of its evolution.

\begin{figure}
\centerline{ \epsfxsize=9cm \epsfbox[200 20 500 250]{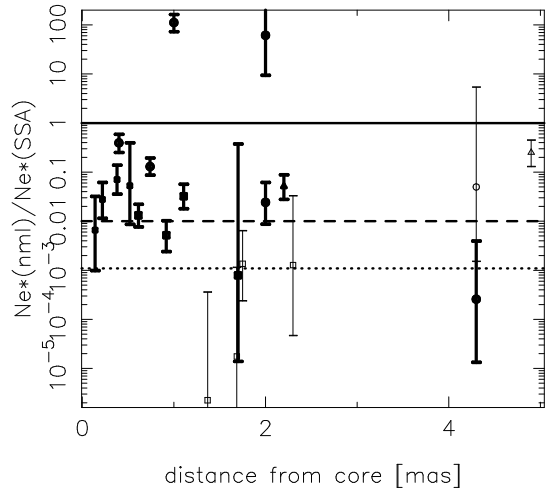} }
\caption{\label{fig:ratio_1}
The ratio $\Nco{}({\rm nml}) / \Nco{}({\rm SSA}) $
as a function of the distance from the core.
Above the solid, horizontal line, the dominance of a normal plasma
is allowed. $\Gamma=15$ is assumed.
        }
\end{figure}

For an assumed Lorentz factor $\Gamma=15$ throughout the jet,
we present the results of $\Nco({\rm nml}) / \Nco{}({\rm SSA})$
as a function of the projected distance
$\rho\sin\varphi$ in figure~\ref{fig:ratio_1}.
The horizontal solid line represents $\Nco=\Nco({\rm nml})$.
It should be noted that $\langle\gamma_-\rangle \sim 10^3$ holds
in equation~(\ref{eq:Ne_nml}) for a normal plasma.
Thus, the solid line gives the upper limit of
$\Nco({\rm nml}) / \Nco{}({\rm SSA})$.
What has be noticed is that 
the possibility of a normal-plasma dominance can be ruled out
if $\Nco({\rm nml}) / \Nco({\rm SSA}) \ll 1$ holds.
There is also depicted a horizontal dashed line representing 
$0.01 \Nco({\rm nml}) / \Nco({\rm SSA}) \approx
 \Nco({\rm pair}) / \Nco({\rm SSA})$
(see eq.~[\ref{eq:DensRat}]).
If the ratio appears near the dashed line,
it indicates that the homogeneous component is pair dominated.
The horizontal dotted line corresponds to 
$(2/1836)\Nco({\rm nml}) / \Nco({\rm SSA})$.
If the ratio appears under this line,
it means that $L_{\rm kin}$ is underestimated 
for that particular component.
If $\nu_{\rm m}$ resides within the observed frequency range,
the point and error bar is indicated by thick pen,
whereas if $\nu_{\rm m}$ 
resides outside of the observed frequency range 
(i.e., if $\nu_{\rm m}$ is extrapolated 
 and hence contains a large error together with $S_{\rm m}$)
it is indicated by thin one.
The two points appearing above the solid line correspond to 
C4 at 1985.8 and 1988.2, which have
the unnatural input parameters as discussed in the foregoing paragraph.

It follows from the figure that 
the upper bound of the $68\%$-error bars for the 11 epochs
appear below $0.1$,
and hence that the jet is likely pair-plasma dominated,
provided $\Gamma \sim 15$.
(C5 at 1984.2 is not counted in the 11 epochs, 
 because it gives too small $\Nco({\rm nml}) / \Nco{}({\rm SSA}) $ value.)
For a smaller $\Gamma$, $\Nco({\rm nml}) / \Nco{}({\rm SSA}) $
further decreases.
However, in this case, the ratio appears lower than the dotted line,
indicating that $L_{\rm kin}$ is underestimated for those
jet components with $\Gamma < 15$,
or that $\langle\gamma_+\rangle \gg 1$ holds for a normal plasma composition.
The ratio shows no evidence of evolution
as a function of the projected distance, $\rho\sin\varphi$, 
under the assumption of a constant $\Gamma$.

It is noteworthy that
$\Nco({\rm nml})/\Nco({\rm SSA})$
is proportional to
$(\gamma_{\rm min}/100)^{-2\alpha-1.56}$
for the spectral index of $-1.15$ for the core.
Therefore, for a typical spectral index $\alpha \sim -0.75$
for the jet components,
the dependence on $\gamma_{\rm min}$ virtually vanishes.
That is, the conclusion does not depend on the assumed value of
$\gamma_{\rm min}\sim 100$ for a normal plasma.

\begin{figure}
\centerline{ \epsfxsize=9cm \epsfbox[200 20 500 250]{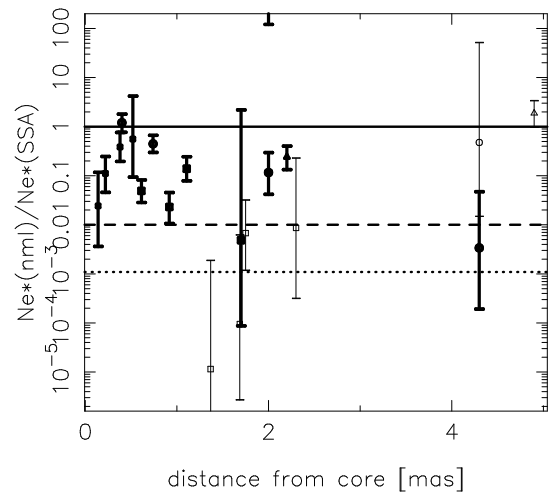} }
\caption{\label{fig:ratio_2}
The same figure as figure~3,
but the different Lorentz factor $\Gamma=20$ is adopted.
        }
\end{figure}

If we assume instead $\Gamma=20$ (rather than $\Gamma=15$)
and $\varphi=2^\circ$,
we obtain the following kinetic luminosity
\begin{equation}
  L_{\rm kin}= 1.3 \times 10^{46} h^{-1.57}
               \langle\gamma_+\rangle
               \left( \frac{\gamma_{\rm min}}{100} \right)^{-1.56}
               K^{0.56}
               \mbox{ergs s}^{-1}.
  \label{eq:Lkin_3C279_4}
\end{equation}
Evaluating the viewing angles of individual components by
equation~(\ref{eq:view_ang}) with $\Gamma=20$,
we can compute $\Nco({\rm nml}) / \Nco({\rm SSA})$ as
presented in figure~\ref{fig:ratio_2}.
It follows from this figure that we cannot rule out the possibility
of a normal-plasma dominance for such large Lorentz factors.
The greater $\Gamma$ we assume,
the greater becomes $\Nco({\rm nml}) / \Nco({\rm SSA})$.

\section{Application to the 3C~279 Jet}    
\label{sec:3C279_appli}

Let us next consider the 3C~279 jet.
This quasar ($z=0.538$)
was the first source found to show superluminal motion
(Cotton et al. 1979; Unwin et al. 1989).
At this redshift, 1 mas ${\rm yr}^{-1}$
corresponds to $\beta_{\rm app}= 28.2 h^{-1}$.

\subsection{Kinetic luminosity}
\label{sec:3C279_Lkin}

In the same manner as in 3C~345 case,
we first consider $\theta_{\rm d, core} \nu / \nu_0$.
Since the turnover frequency of the core is reported to be about
$13$ GHz (Unwin et al. 1989),
we measure $\theta_{\rm d,core}$ at $\nu \leq 10.7$ GHz
(i.e., optically thick frequency range).
At 10.7 GHz,
$\theta_{\rm d,core}$ is reported to be
$0.79$ mas from five-epoch observations by Unwin et al. (1989),
and $0.61$ mas from three-epoch observations by Carrara et al. (1993).
With one-epoch VSOP (VLBI Space Observatory Programme) observation,
$\theta_{\rm d,core}= 1.8 \times 0.28$ mas and
$\theta_{\rm d,core}= 1.8 \times 1.95$ mas were obtained
at 4.8 GHz and 1.6 GHz, respectively (Piner et al. 2000).
As the weighted mean over the 10 epochs, we obtain
\begin{equation}
  \frac{\theta_{\rm d,core} \nu}{\nu_0}= 6.6 \times 10^{-14},
  \label{eq:core_10}
\end{equation}
which leads to
\begin{eqnarray}
  \frac{1}{\nu_0} \frac{\chi \Omega_{r\nu}}{r_1 \sin\varphi}
  &=& 2.8 h^{-1} \left(\frac{\theta_{\rm d,core}}{\rm mas}\right)
                 \frac{\nu}{\nu_\circ}
  \nonumber \\
  &=& 3.7 \times 10^{-13}.
  \label{eq:core_11}
\end{eqnarray}

We next consider the spectral index of the core in the optical thin
frequencies.
From mm to IRAS (90-150 GHz and 270-370 GHz) observations,
Grandi et al. (1996) reported
\begin{equation}
  \alpha= -1.2.
  \label{eq:core_12}
\end{equation}
Adopting this value as the spectral index that
reflects the energy distribution of the power-law electrons in the core,
we obtain
\begin{equation}
  \left( \frac{1}{\nu_0}
         \frac{\chi \Omega_{r\nu}}{r_1 \sin\varphi}
  \right)^{2(5-2\alpha)/(7-2\alpha)}
  = 2.7 \times 10^{-20} h^{-1.57}.
  \label{eq:core_13}
\end{equation}
and
\begin{eqnarray}
  \, & &
  \left[ \pi C(\alpha) \frac{-2\alpha K}{\gamma_{\rm min}{}^{2\alpha+1}}
         \frac{r_1}{r_0}
  \right]^{-4/(7-2\alpha)}
  \nonumber\\
  &=& 1.81 \times 10^{-14} K^{-0.43} \gamma_{\rm min}{}^{-0.56}.
  \label{eq:core_14}
\end{eqnarray}
If we assume a normal plasma dominance, we obtain
\begin{eqnarray}
  L_{\rm kin}
  &=& 3.0 \times 10^{46} h^{-1.57}
      \left(\frac{\gamma_{\rm min}}{100}\right)^{-1.56} K^{0.57}
  \nonumber\\
  &\times& \Gamma(\Gamma-1)
             \left[ \frac{1}{\sin\varphi}
                    \left(\frac{\delta}{1+z}\right)^{2.7}
             \right]^{-0.43}
  \mbox{ergs s}^{-1}.
  \nonumber\\
  \label{eq:Lkin_3C279_0}
\end{eqnarray}

\subsection{Kinematics near the core}
\label{sec:3C279_kinematics}

We now consider $\Gamma$ and $\varphi$ near to the core
to constrain $L_{\rm kin}$ further.
The apparent superluminal velocities of the components in the 3C~279 jet
do not show the evidence of acceleration
and are roughly constant during the propagation.
For example, Carrara et al. (1993) presented
$\beta_{\rm app}= 4.51 h^{-1}$ for C3 during 1981 and 1990 over 15 epochs
at 5, 11, and 22 GHz
and
$\beta_{\rm app}= 4.23 h^{-1}$ for C4 during 1985 and 1990 over 15 epochs
at 11 and 22 GHz.
We thus assume that the apparent motion of the components
represent the fluid bulk motion
and that both $\Gamma$ and $\varphi$ are kept constant
throughout this straight jet.
Since $\beta_{\rm app}$ does not differ very much between C3 and C4,
we adopt $\beta_{\rm app}= 4.51$ as the common value
and apply the same $\Gamma$ and $\varphi$ for these two components.
The errors incurred by the difference in $\beta_{\rm app}$ are small.

The lower bound of $\Gamma$ can be obtained as
$\Gamma > \sqrt{\beta_{\rm app}{}^2 +1}= 4.61$.
The upper bound of $\Gamma$, on the other hand,
can be obtained by the upper bound of $\delta$
for a given $\beta_{\rm app}= 4.51$.
In general, the upper bound of $\delta$ is difficult to infer.
In the case of 3C~279, however,
it is reasonable to suppose that a substantial fraction of
the X-ray flux in the flaring state is produced via synchrotron
self Compton (SSC) process.
Therefore, the Doppler factors
$\sim 3.9$ (Mattox et al. 1993) and
$\sim 5  $ (Henri et al.  1993) derived for the 1991 flare,
and $\sim 6.3$ (Wehrle et al. 1998) for the X-ray flare in 1996,
give good estimates.
We thus consider that $\delta$ does not greatly exceed 10 for
the 3C~279 jet.
Imposing $\delta<10$, we obtain $\Gamma < 6.06$
from $\beta_{\rm app}=4.61$.
Combining with the lower bound,
we can constrain the Lorentz factor in the range
$4.6 < \Gamma < 6.1$.

As case~1, we adopt the smallest bulk Lorentz factor, $\Gamma=4.61$,
which results in $\varphi= 12.5^\circ$ and $\delta= 4.6$
for $\beta_{\rm app}=4.51$.
It is worth comparing this $\delta$ with that measured in the
quiescent state;
Unwin derived $\delta= 3.6 {+4.0 \atop -1.6}$ so that
the calculated X-ray flux may agree with the observation.
Moreover, from the RXTE observation in the 0.1-2.0 keV band
during 1996.06 and 1996.11,
Lawson and M${}^{\rm c}$Hardy (1998)
obtained $2 \mu$Jy in the quiescent state
during 1996.49--1996.54 from 16-epoch observations.
Applying this X-ray flux density to the spectral information
($\nu_{\rm m}$, $S_{\rm m}$, $\alpha$) and $\theta_{\rm d}$
(table~3) for component C4 at epochs 1987.40 and 1989.26
and component C3 at 1984.05,
we obtain $\delta > 1.9$, $>4.0$, and $>3.3$, respectively.
Therefore, we can regard that
the value of $\delta=4.6$ are consistent with the X-ray observations.
In this case, equation~(\ref{eq:Lkin_3C279_0}) gives
\begin{equation}
  L_{\rm kin}
  = 7.2 \times 10^{46} K^{0.57}
    \langle\gamma_+\rangle
    \left(\frac{\gamma_{\rm min}}{100}\right)^{-1.6}
    h^{-1.6}
    \, \mbox{ergs s}^{-1} \, \mbox{(case~1)}.
  \label{eq:Lkin_3C279_1}
\end{equation}

As case~2, we adopt a large Lorentz factor of $\Gamma=6.0$.
In this case, $\beta_{\rm app}= 4.51$ gives
$\delta=9.8$ and $\varphi= 4.45^\circ$.
There is, in fact, another branch of solution
that gives smaller Doppler factor of $2.2$;
however, this solution gives too large
$\Nco({\rm SSA})$'s ($\sim 10^5 \mbox{cm}^{-3}$).
We thus consider only the solution giving $\delta=9.8$ as case~2,
which results in
\begin{equation}
  L_{\rm kin}
  = 3.5 \times 10^{46} K^{0.57}
    \langle\gamma_+\rangle
    \left(\frac{\gamma_{\rm min}}{100}\right)^{-1.6}
    h^{-1.6}
    \, \mbox{ergs s}^{-1} \, \mbox{(case~2)}.
  \label{eq:Lkin_3C279_2}
\end{equation}
We will examine these two cases in the next subsection.

\subsection{Composition of individual components}
\label{sec:3C279_individual}

For the jet component C3,
Unwin et al. (1989) presented $\nu_{\rm m}= 6.8$ GHz,
$S_{\rm m}= 9.4$ Jy, $\alpha= -1.0$, and $\theta_{\rm d}= 0.95$ mas
at epoch 1983.1 (table~3).
It was also reported in their paper that the flux densities were
$5.13 \pm 0.14$ Jy at 5.0 GHz  (at epoch 1984.25),
$4.49 \pm 0.17$ Jy at 10.7 GHz (1984.10), and
$2.58 \pm 0.18$ Jy at 22.2 GHz (1984.09) for C3.
We model-fit the three-frequency data by the function
\begin{equation}
  S_\nu = A_1 \left[ 1-\exp(-A_2 \nu^{\alpha-2.5}) \right].
  \label{eq:model_SSA}
\end{equation}
Assuming $\alpha= -1.0$, we obtain
$\nu_{\rm m}= 6.6$ GHz and $S_{\rm m}= 5.9$ Jy as the best fit
at epoch 1984.10.

For C4, Carrara et al. (1993) presented
$\nu_{\rm m} \sim 11 $ GHz,
$S_{\rm m} \sim 4.3$ Jy, $\alpha= -0.9$, and $\theta_{\rm d} \sim 0.6$ mas
from their 1989-1990 maps at 5, 11, and 22 GHz.
At epoch 1987.4,
we use the flux density of $1.43 \pm 0.17$ Jy at 22 GHz
(Carrara et al. 1993)
and that of $3.95 \pm 0.20$ Jy at 5 GHz (Gabuzda et al. 1999).
We extrapolate the flux density at 11 GHz from those at
1988.17, 1989.26, and 1990.17 presented in Carrara et al. (1993)
to obtain $3.60 \pm 0.20$ Jy.
Assuming $\alpha=-0.9$, we can model fit the three-frequency
data to obtain
$\nu_{\rm m}=6.4$ GHz and $S_{\rm m}=4.4$ Jy.
At this epoch (1987.4),
component C4 is located about 1 mas from the core;
therefore, the angular size vs. distance relation (Carrara et al. 1993)
gives $\theta_{\rm d} \sim 0.6$ mas.
The input parameters for C3 and C4 at these 4 epochs are
summarized in table~3.

In case~1 (see \S~\ref{sec:3C279_kinematics}),
we obtain $\Nco({\rm SSA})= 4.3 \times 10^3$ and $3.0 \times 10^4$,
for C3 at 1983.1 and 1984.10,
and $8.0 \times 10^3$ and $6.4 \times 10^1$
for C4 at 1987.4 and 1989.26, respectively.
These large values of $\Nco({\rm SSA})$ in the first three epochs
result in such (un-physically)
small values of $\Nco({\rm nml})/\Nco({\rm SSA})$ as
$1.4 \times 10^{-4}$,
$2.9 \times 10^{-5}$, and
$1.9 \times 10^{-4}$.
Since even a pair-plasma dominance should be ruled out for
$\Nco({\rm nml})/\Nco({\rm SSA}) \ll 100$ (see eq.~[\ref{eq:DensRat}]),
we consider that the value of $\delta(=4.6)$ is underestimated.

We next examine case~2, in which a larger Doppler factor ($=9.8$)
is adopted.
In this case, we obtain reasonable values of $B$ and $\Nco({\rm SSA})$
as presented in table~3.
It follows that the ratio
$\Nco({\rm nml})/\Nco({\rm SSA})$ becomes
$4.3 \times 10^{-3}$ and $8.7 \times 10^{-3}$,
for C3 at 1983.1 and 1984.10,
and $5.0 \times 10^{-3}$ and $0.62$
for C4 at 1987.4 and 1989.26, respectively.
If $\delta$ exceeds $10$ (or equivalently, if $\Gamma$ exceeds $6$),
the density ratio for C4 at 1989.26 becomes close to unity;
therefore, we cannot rule out the possibility of a normal-plasma
dominance for such large Doppler factors.

On these grounds,
we can conclude that the jet components are dominated
by a pair plasma with $\gamma_{\rm min} \ll 100$,
provided that $\delta<10$ holds,
in the first three epochs.
For the last epoch (1989.26), 
the $68\%$-error bar of $\Nco({\rm nml})/\Nco({\rm SSA})$
appears above $0.1$; 
thus, we cannot rule out the possibility of a normal-plasma dominance.
If $\delta$ is as small as $5$,
not only a normal-plasma dominance but also a pair-plasma dominance
should be ruled out for the first three epochs.
Thus, we consider $\delta$ is greater than $5$ for the 3C 379 jet.

\section{Discussion}
\label{sec:discussion}

In summary, we derived a general scheme to infer the kinetic
luminosity of an AGN jet using a core size observed at a 
single VLBI frequency.
The kinetic luminosity gives an electron density as a function
of the composition of the jet.
If the density deduced under the assumption of a normal-plasma dominance
becomes much less than that obtained independently 
from the theory of synchrotron self-absorption,
we can exclude the possibility of a normal-plasma dominance.
Applying this method to the 3C~345 jet,
we found that components C2, C3, C4, C5, and C7 are 
dominated by a pair plasma with $\gamma_{\rm min} \ll 100$
at 11 epochs out of the total 21 epochs examined,
provided that $\Gamma<15$ holds throughout the jet.
We also investigated the 3C~279 jet
and found that components C3 and C4 are dominated
by a pair plasma with $\gamma_{\rm min} \ll 100$
at three epochs out of the four epochs examined,
provided that $\delta < 10$.

It is noteworthy that the kinetic luminosity computed from
equations~(\ref{eq:Lkin_CPO}) and (\ref{eq:chi_CPO})
has, in fact, weak dependence on the composition.
Substituting $C_{\rm kin}$ for a pair and a normal plasma, we obtain
\begin{eqnarray}
  \frac{L_{\rm kin}({\rm pair})}{L_{\rm kin}({\rm nml})}
  &=&
  \frac{m_{\rm e} \langle\gamma_-({\rm pair})\rangle
                  / \gamma_{\rm min}({\rm pair})}
       {m_{\rm p} / 2\gamma_{\rm min}({\rm nml})}
  \nonumber \\
  &=&
  \frac{1}{9.18}
  \frac{\langle\gamma_-({\rm pair})\rangle}
       {\gamma_{\rm min}({\rm pair})}
  \frac{\gamma_{\rm min}({\rm nml})}
       {100}.
  \label{eq:LkinRat}
\end{eqnarray}
It follows from equation~(\ref{eq:def_gamAVR}) that
$\langle\gamma_-({\rm pair})\rangle / \gamma_{\rm min}({\rm pair}) \sim 3$
holds for $\alpha=-0.75$, for instance.
Thus, we obtain comparable kinetic luminosities irrespectively of
the composition assumed,
if we evaluate them by the method described
in \S~\ref{sec:lineD}.

Let us examine the assumption of $K \sim 0.1$.
For a typical value of $\alpha= -0.75$, we obtain
$(\langle \gamma_- \rangle / \gamma_{\rm min})K
 = (\delta_{\rm eq}/\delta)^{17/2}$,
where $\delta_{\rm eq}$ refers to the
\lq\lq equipartition Doppler factor'' defined by
(Readhead 1994)
\begin{eqnarray}
  \delta_{\rm eq}
  &\equiv&
  \left\{ 10^{-3-6\alpha} F(\alpha)^{34} d(z)^2
  \right.
  \nonumber \\
  &\times&
  \left.  (1+z)^{17-2\alpha}
          \nu_{\rm m}{}^{-35-2\alpha}
          S_{\rm m}{}^{16}
          \theta_{\rm d}{}^{-34}
  \right\}^{1/(13-2\alpha)},
  \nonumber \\
  \label{eq:def_deleq}
\end{eqnarray}
where 
\begin{equation}
  d(z) \equiv \frac{0.65 h}
                   {q_0 z +(q_0-1)(-1+\sqrt{2q_0 z+1})},
  \label{eq:def_dz}
\end{equation}
and $F(\alpha)$ is given in Scott and Readhead (1977);
$\nu_{\rm m}$, $S_{\rm m}$, and $\theta_{\rm d}$ are measured in
GHz, Jy, and mas, respectively.
Therefore, we can expect a rough energy equipartition
between the radiating particles and the magnetic field
if $\delta_{\rm eq}$ is close the $\delta$ derived independently.

Substituting the input parameters presented in table~2,
we can compute $\delta_{\rm eq}$'s for the 3C~345 jet.
The computed $\delta_{\rm eq}$ ranges between
$1.1$ and $67$
except for component C5 at 1984.2, 
which gives $\delta_{\rm eq}=736$ due to unnaturally small $\nu_{\rm m}$.
If we singled out this epoch, we would obtain $\delta_{\rm eq}=23\pm 9$.
The resultant magnetic field strength, $B =6.7 \pm 6.4$ mG, is reasonable.
For the 3C~279 jet,
we obtain $\delta_{\rm eq}=9.5 \pm 1.5$ and
$12 \pm 8$ mG for the four epochs examined.

Since $\delta_{\rm eq}= 23 \pm 9$ for 3C~345
is consistent with those presented in table~2
and since $\delta_{\rm eq}=9.5 \pm 1.5$ for 3C~279
is consistent with those in table~3,
we can expect a rough energy equipartition in these two blazer jets.
However, it may be worth noting that the energy density
of the radiating particles dominates that of the magnetic field
if $\delta$ becomes much smaller than $\delta_{\rm eq}$.
Such a case ($\delta= 4 \sim 12$) was discussed by
Unwin et al. (1992; 1994; 1997) for the 3C~345 jet,
assuming that a significant fraction of the X-rays
was emitted via SSC from the radio-emitting components considered.

We finally discuss the composition variation along the jet.
In the application to the two blazers, we assumed constant
Lorentz factors ($\Gamma$) throughout the jet.
However, $\Gamma$ may in general decrease with increasing
distance from the central engine.
If the decrease of $\Gamma$ is caused by an entrainment
of the ambient matter (e.g., disk wind) consisting of a normal plasma,
it follows from equation~(\ref{eq:Ne_nml}) that
$\Nco({\rm nml})$ and hence $\Nco({\rm nml}) / \Nco({\rm SSA})$
increases at the place where the entrainment occurs.
(Note that the kinetic luminosity will not decrease due to an
entrainment in a stationary jet.)
Therefore, the change of composition from a pair plasma into a normal
plasma along the jet may be found by the method described in this paper.

To study further the change of the composition,
we need detailed spectral information
($\nu_{\rm m}$, $S_{\rm m}$, $\alpha$)
of individual components along the jet.
To constrain the spectral turnover accurately,
we must model-fit and decompose the VLBI images
from high to low frequencies;
in another word, spatial resolution at low frequencies is crucial.
Therefore, simultaneous observations with ground telescopes at higher
frequencies and with space+ground telescopes at lower frequencies
are essential for this study.

\par
\vspace{1pc}\par

\section*{Acknowledgments}

The author wishes to express his gratitude to
Drs. S. Kameno, A.~P. Lobanov and J. Kirk for valuable comments.




\begin{thebibliography}{}
\bibitem[Biretta et al. 1986]{bir86}
    Biretta, J. A., Moore, R. L., \& Cohen, M. H. 1986,
    ApJ 308, 93
\bibitem[Brown et al. 1994]{Bro94}
    Brown, L. F., Roberts, D. H., and Wardle, J. F. C.
    1994, ApJ 437, 108
\bibitem[Celotti and Fabian 1993]{Cel93}
    Celotti, A., Fabian, C. 1993,
    MNRAS 264, 228
\bibitem[Cohen 1985]{Coh85}
    Cohen, M. H. 1985, in Extragalactic Energetic sources,
    Indian Academy of Sciences, Bangalore, p. 1
\bibitem[Despringre and Fraix-Burnet 1997]{des97}
    Despringre, V. \& Fraix-Burnet, D. 1997, AA 320, 26.
\bibitem[Fraix-Burnet 1990]{frai90}
    Fraix-Burnet, D. 1990, A\& A 227, 1
\bibitem[Gabuzda et al. 1999]{gabu99}
    Gabuzda, D. C., Mioduszewski, A. J., Roberts, D. H.,
    Wardle, J. F. C.
    1999, MNRAS 303, 515
\bibitem[Ginzburg et al. 1965]{ginz65}
    Ginzburg, V. L., Syrovatskii, S. I.
    1965, Ann. Rev. Astron. Astrophys. 3, 297
\bibitem[G$\acute{\rm o}$mez et al. 1993] {gom93}
    G$\acute{\rm o}$mez, J. L., Alberdi, A., \& Marcaide, J. M.
    1993, AA 274, 55
\bibitem[G$\acute{\rm o}$mez et al. 1994a] {gom94a}
    G$\acute{\rm o}$mez, J. L., Alberdi, A., \& Marcaide, J. M.
    1994a, AA 284, 51
\bibitem[G$\acute{\rm o}$mez et al. 1994b] {gom94b}
    G$\acute{\rm o}$mez, J. L., Alberdi, A., \& Marcaide, J. M.
    1994b, AA 292, 33
\bibitem[Ghisellini et al. 1992]{ghi92}
    Ghisellini, G., Padovani, P., Celotti, A., \& Maraschi, L. 1992
    MNRAS 258, 776
\bibitem[Gould 1979]{gou79}
    Gould 1979, AA 76, 306
\bibitem[Hardee & Norman 1989]{hard89}
    Hardee, P. E., Norman, M. L. 1989, ApJ 342, 680
\bibitem[Henri et al. 1993]{henr93}
    Henri, G., Pelletier, G., Roland, J.
    1993, ApJ 404, L41
\bibitem[Hewitt et al. 1993]{hewi93}
    Hewitt, A., Burbidge, G. 1993,
    ApJs 87, 451
\bibitem[Hirotani et al. 1999]{hir99}
    Hirotani, K., Iguchi, S., Kimura, M., \& Wajima, K. 1999,
    PASJ 51, 263 (Paper I)
\bibitem[Hirotani et al. 1999]{hir00}
    Hirotani, K., Iguchi, S., Kimura, M., \& Wajima, K. 2000,
    ApJ 545, in press (Paper II)
\bibitem[Impey et al. 1982]{inm82}
    Impey, C. D., Brand, P. W. J. L., Wolstencroft, R. D.,
    and Williams, P. M.
    1982, MNRAS 200, 19
\bibitem[Lawson 1998]{laws96}
    Lawson, A. J., M${}^{\rm c}$Hardy, I. M.
    1998, MNRAS 300, 1023
\bibitem[LeRoux 1961]{lero61}
     Le Roux, E. 1961, Ann. Astrophys. 24, 71
\bibitem[Lobanov & Zensus 1994]{lob94}
    Lobanov, A. P., Zensus, J. A. 1994, in Compact Extragalactic
    Radio Sources, ed. Zensus, J. A. , Kellermann, K. I.,
    National Radio Astronomy Observatory Workshop No. 23,
    p. 157
\bibitem[Lobanov 1998]{lob98}
    Lobanov, A. P. 1998, AA 330, 79
\bibitem[Lobanov & Zensus 1999]{lob99}
    Lobanov, A. P., Zensus, J. A. 1999,
    ApJ 521, 509
\bibitem[Marscher 1983]{mar83}
    Marscher, A. P. 1983, ApJ 264, 296
\bibitem[Moshir et al. 1990]{mos90}
    Moshir, M., Kopan, G., Conrow, T., McCallon, H., Hacking, P.,
    Gregorich, D., Rohrbach, G., Melnyk, M., Rice, W., Fullmer, L. et al.
    1990, in Infrared Astronomical Satellite Catalogs,
    the faint source catalog, ver. 2.0
\bibitem[Mattox et al. 1993]{matt82}
    Mattox, J. R., Bertsch, D. L., Chiang, J. et al.
    1993, ApJ 410, 609
\bibitem[Neugebauer 1982]{neu82}
    Neugebauer, G., Soifer, B. T., Matthews, K., Margon, B.,
    Chanan, G. A. AJ 1982, 87, 1639
\bibitem[Pauliny-Toth et al. 1981]{pau81}
    Pauliny-Toth I. I. K., Preuss E., Witzel A.,
    Graham D., Kellermann I. I., Ronnang B. 1981,
    AJ 86, 371
\bibitem[Pelletier \& Marcowith 1998]{pel98}
    Pelletier, G., \& Marcowith, I. 1998, ApJ 502, 598
\bibitem[Pelletier \& Roland 1989]{pel89}
    Pelletier, G., Roland, J. 1989, A \& A 224, 24
\bibitem[Pelletier \& Sol 1992]{pel92}
    Pelletier, G., \& Sol, H. 1992, MNRAS 254, 635
\bibitem[Piner et al. 2000]{pine00}
    Piner, B. G., Edwards, P. G., Wehrle, A. E., Hirabayashi, H.,
    Lovell, J. E. J., and Unwin, S. C. 2000,
    ApJ 537, 91 
\bibitem[Qian et al. 1996]{qian95}
    Qian, S. J., Krichbaum, T. P., Zensus, J. A., Steffen, W., Witzel, A.
    1996, A \& A 308, 395
\bibitem[Readhead 1994]{rea94}
    Readhead 1994,
    ApJ 426, 51
\bibitem[Reynolds et al. 1996]{rey96}
    Reynolds, C. S., Fabian, A. C., Celotti, A., \& Rees, M. J. 1996
    MNRAS 283, 873
\bibitem[Ros et al. 2000]{ros00}
    Ros, E., Zensus, J. A., Lobanov, A. P. 2000, A \& A 354, 55
\bibitem[Scott and Readhead 1977]{sco77}
    Scott, M. A., Readhead, A. C. S. 1977,
    MNRAS 180, 539
\bibitem[Sitko et al. 1982]{Sit82}
    Sitko, M. L. Stein, W. A., Zhang, Y. X., Wisniewski, W. Z.
    1982, ApJ 259, 486
\bibitem[Sol et al. 1989]{sol89}
    Sol, H. Pelletier, G., \& Ass$\acute{\rm e}$o, E.
    1989, MNRAS 237, 411
\bibitem[Steffen et al. 1995]{stef95}
    Steffen, W., Zensus, J. A., Krichbaum, T. P., Witzel, A.,
    Qian, S. J. 1995, A \& A 302, 335
\bibitem[Unwin et al. 1989]{unw89}
    Unwin, S. C., Cohen, M. H., Biretta, J. A.,
    Hodges, M. W., and Zensus, J. A.
    1989, ApJ 340, 117
\bibitem[Unwin et al. 1994]{unw92}
    Unwin, S. C., Wehrle, A. E.
    1992, ApJ 398, 74
\bibitem[Unwin et al. 1994]{unw94}
    Unwin, S. C., Wehrle, A. E., Urry C. M., Gilmore, D. M.,
    Barton, E. J., Kjerulf, B. C., Zensus, J. A., and
    Rabaca, C. R. 1994, ApJ 432, 103
\bibitem[Unwin et al. 1997]{unw97}
    Unwin, S. C., Wehrle, A. E., Lobanov, A. P., Zensus, J. A.,
    \& Madejski, G. M. 1997,
    ApJ 480, 596
\bibitem[Wardle et al. 1998]{ward98}
    Wardle, J. F. C., Homan, D. C., Ojha, R, Roberts, D. H.
    Nature 395, 457
\bibitem[Wehrle et al. 1998]{wehr90}
    Wehrle, A. E., Pian E., Urry, C. M., et al.
    1998, ApJ 497, 178
\bibitem[Zensus et al. 1995]{zen95}
    Zensus, J. A., Cohen, M. H., \& Unwin, S. C.  1995,
    ApJ 443, 35
\bibitem[Zensus et al. 1997]{zen97}
    Zensus, J. A.  1997,
    Ann. Rev. Astron. Astrophys. 35, 607
\end{thebibliography}
\end{document}